\documentclass[12pt,letterpaper]{JHEP3}

\usepackage{amsmath}
\usepackage{cancel}
\usepackage{epsfig}
\usepackage[vcentermath]{syoungtab}
\usepackage{lscape}
\usepackage{tabularx}

\newcommand{\ayng}[2]{\overbrace{\yng(#2,#2)\raise 3.2pt\hbox{\,$\cdots$\yng(1)}}^{#1}}

\newcommand{\Vd}[3]{{\cal V}^{#3}_{(#1)[#2]}}
\newcommand{\Vdb}[3]{{\overline{\cal V}}^{#3}_{(#1)[#2]}}
\newcommand{\cv}{{\cal V}}
\newcommand{\ocv}{{\overline{\cal V}}}

\preprint{Brown-HET-1589}

\title{\Large Two-Loop Spectroscopy of Short ABJM Operators}

\author{Georgios Papathanasiou and Marcus Spradlin\\
Department of Physics\\
Brown University, Providence, Rhode Island 02912, USA\\
E-mail:
\email{Georgios\_Papathanasiou@brown.edu},
\email{Marcus\_Spradlin@brown.edu}}

\abstract{We study the spectrum of anomalous dimensions of short operators in planar ABJM theory at two loops. Specifically we develop a method for solving the $OSp(6|4)$ Bethe ansatz equations for a certain class of unpaired length-4 states with arbitrarily high number of excitations, and apply it to identify three new sequences of rational eigenvalues. Results for low-lying paired states in the $OSp(4|2)$ sector are obtained by direct diagonalization of the spin chain Hamiltonian. We also study the $SL(2|1)$ sector and identify the set of states that corresponds to the $SL(2)$-like Bethe ansatz of Gromov and Vieira. Finally we extend part of our analysis to length-6 operators.}

\begin{document}

\section{Introduction}

The conformal ${\cal{N}} = 6$ supersymmetric Chern-Simons matter theory proposed by Aharony, Bergman, Jafferis and Maldacena (ABJM)~\cite{Aharony:2008ug} in order to describe the worldvolume of M2-branes (building on the earlier work~\cite{Schwarz:2004yj,Gustavsson:2007vu,Bagger:2007jr}) has been the subject of intense study.  It is a 3-dimensional theory with gauge group $U(N) \times U(N)$ and superconformal group $OSp(6|4)$ (see also~\cite{Gaiotto:2007qi,Benna:2008zy}).

The dynamical fields of the ABJM theory are four complex scalars $\phi_i$ and their fermionic partners $\bar\psi^i$, $i=1,...,4$, which transform in the $(N,\overline{N})$ of the gauge group and in the so-called `singleton' representation of the superconformal group, denoted $\cv^1$. In addition we have their complex conjugate fields $\bar\phi^i, \psi_i$ in the $(\overline{N},N)$ and conjugate singleton $\ocv^1$ representations of the gauge and superconformal groups respectively. Finally the gauge fields are non-dynamical, with Chern-Simons actions and opposite levels $+k$ and $-k$ for the two gauge groups.
Gauge invariant single-trace operators are formed by taking the trace
of a product of an even number of matter fields (or their covariant
derivatives), alternating between $\cv^1$ and $\ocv^1$.

Similarly to the 't Hooft limit in ${\cal{N}} = 4$ Yang-Mills (SYM) theory, the existence of two parameters $N,k$ allows us to take them to infinity with $\lambda=N/k$ fixed. In this limit the theory admits a dual description in terms of type IIA string theory on $AdS_4\times {\mathbb{P}}^3$ and provides an important example of the $AdS_4/CFT_3$ correspondence~\cite{Aharony:2008ug}.

Integrability plays a key role in unveiling the structure of planar ABJM theory~\cite{Minahan:2008hf,Gaiotto:2008cg,Bak:2008cp,Spill:2008yr,Kristjansen:2008ib,Gromov:2008qe,Ahn:2008aa,Gromov:2009tv}. At leading order in the weak coupling expansion, it was proven that the spectrum of of anomalous dimensions for certain subsets of single-trace, gauge invariant operators of the theory is encoded in an integrable spin chain Hamiltonian~\cite{Minahan:2008hf,Gaiotto:2008cg,Bak:2008cp}, of the general form
\begin{equation*}
\Delta_2  = \lambda^2 \sum_{i=1}^{2 L} (D_2)_{i,i+1,i+2}\,
\end{equation*}
where the Hamiltonian density $D_2$ acts simultaneously on three adjacent sites of a spin chain with $2L$ sites in total. This result was extended with the construction of the full 2-loop dilatation operator~\cite{Zwiebel:2009vb,Minahan:2009te}, and aspects related to integrability have been studied also at higher loops~\cite{Bak:2009mq,Minahan:2009aq,Bak:2009tq}.

A set of all-loop Bethe equations encoding the full asymptotic spectrum of the planar theory has been proposed in~\cite{Gromov:2008qe}, with a later proposal~\cite{Gromov:2009tv} also incorporating corrections due to wrapping interactions. The all-loop equations were in turn derived from a conjectured S-matrix~\cite{Ahn:2008aa}, which successfully passed a 2-loop test \cite{Ahn:2009zg}. An unusual feature of the proposed S-matrix is that that the scattering between odd- and even-site excitations is reflectionless, studied in more detail in~\cite{Ahn:2009tj}.

In this paper we study the spectrum of the 2-loop dilatation operator of the ABJM theory for states of length 4 and 6 in various sectors of the superconformal group $OSp(6|4)$, using a combination of Bethe ansatz techniques and direct Hamiltonian diagonalization.
Specifically we present
new analytic formulas for
three new infinite sequences of rational eigenvalues, and a numerical method
for determining further (irrational) eigenvalues with relative ease.
Our results for length-4
states are summarized in Table~\ref{tab4}, which may be thought of
as the ABJM theory analogue of Table 3 of~\cite{Beisert:2004di} (also
Table 3.10 of~\cite{Beisert:2004ry}) for ${\cal{N}}=4$ SYM.
See~\cite{Beccaria:2009ny,Beccaria:2009wb} for eigenvalues of other
states in the ABJM theory.

In section 2 we review a few necessary
details regarding $OSp(6|4)$ representation theory
(mostly from~\cite{Papathanasiou:2009en}).
We begin section 3 by presenting our results for the eigenvalue
sequences, then describe the special states which have the
energies shown in Table~\ref{tab4}, explain how the results
were obtained from the Bethe ansatz equations (BAE), and finally
describe a numerical method which may be used to find more general
eigenvalues.
Section 4 explains an apparent `coincidence' in the sequence of
eigenvalues as a consequence of two short multiplets of $OSp(6|4)$
combining into a long multiplet at finite coupling.  In section 5 we
discuss various subsectors in detail (including some results
from direct Hamiltonian diagonalization in the $OSp(4|2)$ sector) before
mentioning a few comments on length-6 operators in section 6.

\section{Preliminaries}

The superconformal group of the ABJM theory, $OSp(6|4)$, has $Sp(4,\mathbb{R})\times SO(6)$ as its bosonic subgroup, where $Sp(4,\mathbb{R})\simeq SO(2,3) $ is the conformal group of 3-dimensional spacetime and $SO(6)\simeq SU(4)$ is the $R$-symmetry group. Representations of $OSp(6|4)$ formed by any number $f$ of fundamental fields of the theory (and their derivatives) are conventionally labeled by the Cartan charges $[\Delta,j;d_1,d_2,d_3]$ and the length $f$, where $(\Delta,j)$ are the classical scaling dimension and spin charges of $SO(2,3)$ and $[d_1,d_2,d_3]$ are the Dynkin labels of  $SU(4)$.

Alternatively, we can also characterize $OSp(6|4)$ representations with a set of super-Young tableau (SYT) labels $(k_1,k_2,\ldots,k_n)$, where the number of labels $n$ can take various values, but for our discussion won't extend beyond $n=3$. The interested reader can find out more about the physical significance of the SYT and their corresponding labels in the context of the ABJM theory
in~\cite{Papathanasiou:2009en}
(see~\cite{Baha Balantekin:1980qy,Baha Balantekin:1980pp,Baha Balantekin:1981bk,Bars:1982se,Morel:1984vw,Hurni:1985vk} for a more general discussion). Here we will use the SYT labels as a significantly more compact formalism for describing multiplets than the Cartan charges. At any stage one can translate from the SYT labels to the usual language of the Cartan charges with the help of the relations
\begin{equation}
\begin{aligned}
\Delta&=\frac{1}{2}\left(\max(k_1-3,0)+\max(k_2-3,0)+f \right)\,,\\
j&=\frac{1}{2}(\max(k_1-3,0)-\max(k_2-3,0))\,,\\
d_1&=f-\sum_{i=1}^n \min(k_i,2)\,, \qquad
d_2 = \sum_{i=1}^n \delta_{k_i,2}\,, \qquad
d_3 = \sum_{i=1}^n \delta_{k_i,1}\,.
\end{aligned}
\label{yt2dynkin}
\end{equation}
We will denote a multiplet in the Cartan or SYT formalism as $\Vd{\Delta, j}{d_3,d_2,d_1}{f}$ or $\cv^f_{k_1,k_2,\ldots,k_n}$ respectively, and we will use
a bar to denote the `conjugate' of a representation, which is
obtained by reversing the order of the $SU(4)$ labels, $\Vdb{\Delta,j}{d_1,d_2,d_3}{f}=\Vd{\Delta, j}{d_3,d_2,d_1}{f}$.
The complete set of length-4 multiplets, which will be the focus of most
of our study, are displayed in Table~\ref{f=4reps}.

\TABLE[t]{
\begin{tabular}{|clccccl|} \hline
&&&&&& \\
$1$ & $\cv^4 = \Vd{2,0}{4,0,0}{}$ &&&& $\syng(1,1,1,1)$ & $\cv^4_{1,1,1,1} = \Vd{2,0}{0,0,4}{4} = \ocv^4$ \\ [12pt]
$\syng(1)$ & $\cv^4_1 = \Vd{2,0}{3,0,1}{4}$ &&&& $\syng(1,1,1)$ & $\cv^4_{1,1,1} = \Vd{2,0}{1,0,3}{2} = \ocv^4_1$ \\ [12pt]
$\syng(2)$ & $\cv^4_{2} = \Vd{2,0}{2,1,0}{4}$ &&&& $\syng(2,1,1)$ & $\cv^4_{2,1,1} = \Vd{2,0}{0,1,2}{4} = \ocv^4_2$ \\ [12pt]
$\overbrace{\styng(2)\cdots\styng(1)}^{k\ge\,3}$ & $\cv^4_k = \Vd{\frac{k+1}{2},\frac{k-3}{2}}{2,0,0}{4}$ &&&& $\overbrace{\styng(2,1,1)\raise 6.7pt\hbox{\,$\cdots$\styng(1)}}^{k\ge\,3}$
& $\cv^4_{k,1,1} = \Vd{\frac{k+1}{2},\frac{k-3}{2}}{0,0,2}{4} = \ocv^4_k$ \\ [12pt]
$\syng(1,1)$ & $\cv^4_{1,1} = \Vd{2,0}{2,0,2}{4}$ &&&& $\overbrace{\styng(2,1)\raise 3.2pt\hbox{\,$\cdots$\styng(1)}}^{k\ge\,3}$
& $\cv^4_{k,1} = \Vd{\frac{k+1}{2},\frac{k-3}{2}}{1,0,1}{4}$ \\ [12pt]
$\syng(2,1)$ & $\cv^4_{2,1} = \Vd{2,0}{1,1,1}{4}$ &&&& $\overbrace{\styng(2,2)\raise 3.2pt\hbox{\,$\cdots$\styng(1)}}^{k\ge\,3}$
& $\cv^4_{k,2} = \Vd{\frac{k+1}{2},\frac{k-3}{2}}{0,1,0}{4}$ \\ [12pt]
$\syng(2,2)$ & $\cv^4_{2,2} = \Vd{2,0}{0,2,0}{4}$ &&&& $\overbrace{\underbrace{\styng(1,1)\cdots\styng(1,1)}_{k_2\ge 3}\raise 3.2pt\hbox{\,$\cdots$\styng(1)}}^{k_1\ge k_2}$
& $\cv^4_{k_1,k_2} = \Vd{\frac{k_1+k_2-2}{2},\frac{k_1-k_2}{2}}{0,0,0}{4}$ \\ [12pt] \hline
\end{tabular}
\caption{The $f=4$ $OSp(6|4)$ multiplets.}
\label{f=4reps}
}

The decomposition of all states of the ABJM theory with $f=4$ into irreducible $OSp(6|4)$ multiplets is
\begin{multline}
(\cv^1\otimes\ocv^1)^2= \sum_{j=1}^\infty \frac{1}{2} j(j+1)\left(\cv^4_{2j}+\ocv^4_{2j}\right) \\
+ \sum_{j=1}^\infty \sum_{p=1}^j \left\{\left[j(j+1)-p^2\right] \cv^4_{2j,2p}+\left[j^2-p(p-1)\right]\cv^4_{2j-1,2p-1}\right\}.
\label{f=4decomp1}
\end{multline}
Perhaps more important is the subset of multiplets which are (graded) symmetric under exchange of the two $\cv^1\otimes\ocv^1$ factors, since these correspond to physical gauge invariant operators,
\begin{equation}
\begin{aligned}
(\cv^1\otimes\ocv^1)^2_+&=\sum_{j=1}^\infty j(j+1)(\cv^4_{4j}+\ocv^4_{4j}+\cv^4_{4j+2}+\ocv^4_{4j+2})+[2j(j-1)+1]\cv^4_{4j-3,1} \\
&\quad + \sum_{j=1}^\infty \sum_{p=1}^j
\left\{2\left[j(j+1)-p^2\right](\cv^4_{4j,4p}+\cv^4_{4j+2,4p})\right. \\
&\quad\quad\qquad\qquad +2\left[j^2-p(p-1)\right](\cv^4_{4j-1,4p-1}+\cv^4_{4j-1,4p-3}) \\
&\quad\quad\qquad\qquad+\left[2j^2-1-2p(p-1)\right] (\cv^4_{4j-2,4p-2}+\cv^4_{4j,4p-2}) \\
&\quad\quad\qquad\qquad\left.+ \left[2j(j+1)+1-2p^2\right] (\cv^4_{4j+1,4p-1}+\cv^4_{4j+1,4p+1}) \right\}.
\end{aligned}
\label{f=4sym_decomp}
\end{equation}
We should note that the corresponding formulas for the $OSp(4|2)$ sector are given by the same expressions if we simply drop all multiplets that have a second SYT label greater or equal to 3.

\section{Unpaired $OSp(6|4)$ Multiplets}\label{chapter_unpaired_multiplets}

We have developed a method for solving the 2-loop $OSp(6|4)$ BAE for certain states of each irreducible representation appearing in the cyclic tensor product decomposition (\ref{f=4sym_decomp}). In this manner, we have been able to compute the low-lying spectrum of 2-loop anomalous dimensions $\Delta_2$, and consequently identify three new infinite sequences of rational eigenvalues.

Our results are summarized in Table~\ref{tab4}, with the new sequences given by\footnote{Since we are solely interested in the 2-loop contribution to the anomalous dimension, we omit the overall $\lambda^2$ factor in all formulas for scaling dimensions.}
\begin{equation}\label{new_eigenvalue_sequences}
\begin{aligned}
\Delta_2&=4\big(S_1(m)+S_{-1}(m)\big)+\frac{2(1-(-1)^m)}{m+1}+8 \qquad
\qquad\qquad~~\mbox{for $\cv^4_{2m+1,3}$},\\
\Delta_2&=4\big(S_1(m)-S_{-1}(m)\big)+\frac{1+(-1)^m}{m+1}-\frac{1-(-1)^m}{m}+4
\qquad \mbox{for $\cv^4_{2m,4}$},\\
\Delta_2&=4\big(S_1(m)+S_{-1}(m)\big)+\frac{2(1-(-1)^m)}{m+1}+\frac{32}{3}
\qquad\qquad\qquad\mbox{for $\cv^4_{2m+1,5}$ ($m$ odd)}.\\
\end{aligned}
\end{equation}
In these expressions $S_a(j)$ is a generalized harmonic number, defined as
\begin{equation}
S_a(j)\equiv\sum_{i=1}^j \frac{(\mbox{sign}\,a)^i }{i^{|a|}}\,,
\end{equation}
with $S_1(j)$ corresponding to the ordinary harmonic numbers.

Throughout this section we work with the Bethe ansatz of the ABJM spin chain corresponding to the distinguished Dyn\-kin diagram of Figure~\ref{DynkinDiagram},
\begin{eqnarray}
 \label{betheeq}
\left(\frac{u_i+i/2}{u_i-i/2}\right)^L&=&\prod_{k=1,k\ne j}^{M_u}\frac{u_i-u_k+i}{u_i-u_k-i} \prod_{k=1}^{M_r}\frac{u_i-r_k-i/2}{u_i-r_k+i/2}\nonumber\\
\left(\frac{v_i+i/2}{v_i-i/2}\right)^L&=&\prod_{k=1,k\ne i}^{M_v}\frac{v_i-v_k+i}{v_i-v_k-i}
\prod_{k=1}^{M_r}\frac{v_i-r_k-i/2}{v_i-r_k+i/2}\nonumber\\
1&=& \prod_{k=1,k\ne i}^{M_r}\frac{r_i-r_k+i}{r_i-r_k-i}\prod_{k=1}^{M_u}\frac{r_i-u_k-i/2}{r_i-u_k+i/2}\prod_{k=1}^{M_v}\frac{r_i-v_k-i/2}{r_i-v_k+i/2}\prod_{k=1}^{M_s}\frac{r_i-s_k-i/2}{r_i-s_k+i/2}\nonumber\\
1&=&\;\:\,\prod_{k=1}^{M_r}\frac{s_i-r_k-i/2}{s_i-r_k+i/2}\prod_{k=1}^{M_w}\frac{s_i-w_k+i/2}{s_i-w_k-i/2}\nonumber\\
1&=&\prod_{k=1,k\ne i}^{M_w}\frac{w_i-w_k-i}{w_i-w_k+i}\prod_{k=1}^{M_s}\frac{w_i-s_k+i/2}{w_i-s_k-i/2}\,
\end{eqnarray}
as derived
in~\cite{Minahan:2008hf}.

Acceptable solutions to these equations cannot have two or more roots of the same kind being equal, and for physical states we additionally need to impose the trace cyclicity condition,
\begin{equation}\label{cyclicity}
\prod_{k=1}^{M_u}\frac{u_k+i/2}{u_k-i/2}\prod_{k=1}^{M_v}\frac{v_k+i/2}{v_k-i/2}=1\,.
\end{equation}
The 2-loop anomalous dimensions, or energies, of the states are given by
\begin{equation}\label{energies}
\Delta_2=\sum_{k=1}^{M_u}\frac{1}{u^2_k+\frac{1}{4}}+\sum_{k=1}^{M_v}\frac{1}{v^2_k+\frac{1}{4}}\,.
\end{equation}

\FIGURE{
\includegraphics[height=4cm]{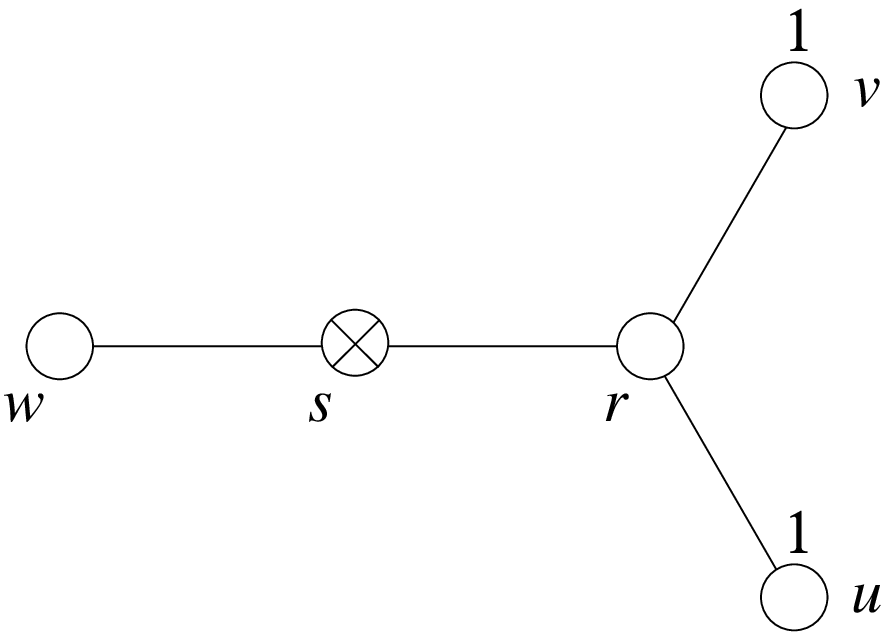}
\caption{The $OSp(6|4)$ distinguished Dynkin diagram. The Bethe roots
$s_i$, $w_i$, $r_i$, $v_i$
and $u_i$ in~(\ref{betheeq}) correspond respectively to excitations of the
indicated simple roots.}
\label{DynkinDiagram}
}

Evidently the Bethe equations~(\ref{betheeq}) are invariant when changing the sign of every root, so their solutions will generically come in pairs related by this transformation, which due to~(\ref{energies}) will have the same energies. The only case when this arrangement of solutions into pairs doesn't occur is for solutions that are themselves invariant under this transformation\footnote{Solutions only differing in the ordering of roots are of course equivalent.}, namely if $u_{i+M_u-1}=-u_i$ and similarly for the other four kinds of roots. Following the analysis for the BAE of ${\cal{N}}=4$
SYM~\cite{Beisert:2004ry} we refer to the latter kind of states as ``unpaired''.

When $M_u=M_v$ the Bethe equations are also symmetric
under the map $u_i\leftrightarrow v_i$, so a similar pairing of states should occur, except for solutions that obey $u_i=v_i$. Indeed such states can be found when examining length-4 multiplets, as apart from $\cv^4_k$ (and its conjugate) all other multiplets do have $M_u=M_v$. Hence most of our analysis will focus on solutions to the Bethe equations that are both unpaired and obey $u_i=v_i$, as is the case for all solutions shown in Table~\ref{tab4}. Such solutions will have multiplicity 1 by construction, as they are mapped to themselves under the two aforementioned symmetry transformations.

\TABLE[t]{
\begin{tabularx}{\textwidth}{
|>{\hsize=0.55\hsize\centering\arraybackslash}X
|>{\hsize=1.05\hsize\centering\arraybackslash}X
|>{\hsize=1.05\hsize\centering\arraybackslash}X
|>{\hsize=1.05\hsize\centering\arraybackslash}X
|>{\hsize=1.05\hsize\centering\arraybackslash}X
|>{\hsize=1.05\hsize\centering\arraybackslash}X
|>{\hsize=1.05\hsize\centering\arraybackslash}X
|>{\hsize=1.05\hsize\centering\arraybackslash}X
|>{\hsize=1.05\hsize\centering\arraybackslash}X
|>{\hsize=1.05\hsize\centering\arraybackslash}X|}\hline
$j \backslash p$&1&2&3&4&5&6&7&8&9\\\hline
1 & 0& & & & & & & &\\\hline
2 & &8 & & & & & & &\\\hline
3 & & &10& & & & & &\\\hline
4 & &8 & &$\frac{38}{3}$& & & & &\\\hline
5 & 4& &12& & 14.43& & & &\\\hline
6 & &$\frac{32}{3}$ & &14& &15.94& & &\\\hline
7& & &13& &$\frac{47}{3}$& &17.20 & &\\\hline
8& &$\frac{32}{3}$ & &$\frac{226}{15}$&&17.01& &18.29&\\\hline
9& 6& &14 &&16.59& &18.15& &19.25\\\hline
10& &$\frac{184}{15}$ & &$\frac{238}{15}$& &17.84& &19.14&\\\hline
11& & &$\frac{44}{3}$& &$\frac{52}{3}$& &18.91 & &20.02\\\hline
12& &$\frac{184}{15}$ & &$\frac{1738}{105}$& &18.52& &19.84 &\\\hline
13& $\frac{22}{3}$& &$\frac{46}{3}$& &17.96& &19.54 & & $\ddots$ \\\hline
14& &$\frac{1408}{105}$ & &$\frac{1798}{105}$& &19.11 & & $\ddots$ & \\\hline
15& & &$\frac{95}{6}$& &$\frac{37}{2}$ & & $\ddots$ & & $\ddots$ \\\hline
16& &$\frac{1408}{105}$ & &$\frac{5554}{315}$ &&$\ddots$ &&$\ddots$ & \\\hline
17& $\frac{25}{3}$& &$\frac{49}{3}$ & & $\ddots$ &&$\ddots$ &&$\ddots$ \\\hline
\end{tabularx}
\caption{Unpaired (multiplicity 1) eigenvalues of the multiplets denoted by $\cv^4_{j,p}$ in SYT
notation. The numbers containing decimal points are numerical approximations
to the exact values, which
are irrational.
The first column was derived
in~\cite{Beccaria:2009ny}.
The interested reader may find analogous results for ${\cal{N}}=4$ SYM
in Table 3 of~\cite{Beisert:2004di}.
}\label{tab4}
}

Finally, in order to find the BAE that correspond to a certain multiplet we need the relation between the root excitation numbers $M_u,\ldots,M_w$ of the former and the Cartan charges of the latter, which are given
by~\cite{Minahan:2009te,Papathanasiou:2009en}
\begin{equation}\label{excitations_to_cartan}
\left(\begin{array}{l}
M_u\\
M_v\\
M_r\\
M_s\\
M_w
\end{array}\right)=\left(\begin{array}{l}
\phantom{2} \Delta\phantom{\mathord{}-2L}-\frac{3}{4}d_1-\frac{1}{2}d_2-\frac{1}{4}d_3\\
\phantom{2} \Delta\phantom{\mathord{}-2L}-\frac{1}{4}d_1-\frac{1}{2}d_2-\frac{3}{4}d_3\\
2\Delta-\phantom{2}L-\frac{1}{2}d_1-\phantom{\frac{1}{4}}d_2-\frac{1}{2}d_3\\
2\Delta-2L\\
\phantom{2}\Delta-\phantom{2}L-j\\
\end{array}\right),
\end{equation}
where $L=\frac{f}{2}$ is half the length of the spin chain or multiplet.

In the following sections we explain how the new results of Table~\ref{tab4} and equation~(\ref{new_eigenvalue_sequences}) were obtained.

\subsection{Proof of the $\cv^4_{2m,2}$ Eigenvalue Sequence}\label{V2m}

Let us start with the $\cv^4_{2m,2}$ multiplets occupying the second column of Table~\ref{tab4}, as the $\cv^4_{4m+1,1}$ multiplets in the first column correspond to the twist-2 states whose anomalous dimensions have been already
determined
in~\cite{Beccaria:2009ny},
\begin{equation}\label{V4m+1,1}
\begin{aligned}
\Delta_2&=4 S_1(m)
\qquad \mbox{for $\cv^4_{4m+1,1}$}, \\
\end{aligned}
\end{equation}
and were merely included in the table for completeness.

Using~(\ref{yt2dynkin}) and~(\ref{excitations_to_cartan}) we find that the root excitation numbers of $\cv^4_{2m,2}$ are
\begin{equation}
[M_u,M_v,M_r,M_s,M_w]=[m,m,2m-2,2m-3,0]
\end{equation}
for $m\ge2$. The Bethe equations for the $s$ roots then take the form
\begin{equation}\label{sBAE}
\prod_{k=1}^{2m-2}\frac{s_i-r_k-i/2}{s_i-r_k+i/2}=1\,, \qquad i=1,2,\ldots,2m-3,
\end{equation}
and it is easy to show that each of them becomes the same polynomial equation of order $2m-3$ for the single variable $s_i$. Since the number of $s$ roots is also $2m-3$, they precisely correspond to the solutions of this single polynomial equation.

With the help of the argument of appendix~\ref{useful_theorem} for $n\to2m-2$, $x_i\to s_i$, $y_i\to r_i$ and $a\to \mp i/2$, it is then easy to prove that~(\ref{sBAE}) implies the relation\footnote{
This kind of replacement is the simplest case of a more general procedure for exchanging one set of supersymmetric BAE for another, called fermionic root dualization. In the appendix A we review the aspects which are relevant to our discussion,
see~\cite{Beisert:2005di} for more information.}
\begin{equation}
\prod_{k=1}^{2m-3}\frac{r_i-s_k-i/2}{r_i-s_k+i/2}=\prod_{k=1,k\ne i}^{2m-2}\frac{r_i-r_k-i}{r_i-r_k+i}\,.
\end{equation}
Plugging this relation into the Bethe equations for the $r$ roots leads
to the greatly simplified equation
\begin{equation}\label{rBAE}
\prod_{k=1}^{m}\frac{r_i-u_k-i/2}{r_i-u_k+i/2}\prod_{k=1}^{m}\frac{r_i-v_k-i/2}{r_i-v_k+i/2}=1\,.
\end{equation}
This equation is similar to~(\ref{sBAE}) and we could repeat the same story here, except for a subtlety which arises. The equation turns into a polynomial of degree $2m-1$ in $r_i$, but there are only $2m-2$ different $r_i$, so the polynomial equation also gives one additional, ``dual'' root. In order to overcome this difficulty we first notice that for $r_i\to0$, the equation~(\ref{rBAE}) turns into the cyclicity condition~(\ref{cyclicity}), and hence for cyclically invariant states it is a solution. Then we can ensure that $r_i=0$ corresponds to the additional, dual root, by restricting to unpaired states (which means $r_{M_r+1-i}=-r_i$, and similarly for the other roots), since $M_r=2m-2$ is even and we can't have any two roots coinciding. Thus, for unpaired states, application of the argument of appendix~\ref{useful_theorem} yields the relation
\begin{equation}\label{from_rBAE}
\prod_{k=1}^{2m-2}\frac{u_i-r_k-i/2}{u_i-r_k+i/2}=\frac{u_i+i/2}{u_i-i/2}\prod_{k=1,k\ne i}^{m}\frac{u_i-u_k-i}{u_i-u_k+i}\prod_{k=1}^{m}\frac{u_i-v_k-i}{u_i-v_k+i}
\end{equation}
and similarly for $u_j\leftrightarrow v_j$.
If we further substitute~(\ref{from_rBAE}) into the Bethe equations for the $u$ and $v$ roots, these simplify drastically to
\begin{equation}\label{sl(2|1)_for_L=1}
\begin{aligned}
\frac{u_i+i/2}{u_i-i/2}=\prod_{k=1}^{m}\frac{u_i-v_k-i}{u_i-v_k+i}\,,\\
\frac{v_i+i/2}{v_i-i/2}=\prod_{k=1}^{m}\frac{v_i-u_k-i}{v_i-u_k+i}\,,
\end{aligned}
\end{equation}
which we recognize as the $SL(2|1)$ BAE for length-2 ($L=1$)
operators~\cite{Minahan:2009te}, corresponding to the $\cv^2_{2m+1}$ multiplets\footnote{The connection between the notation
of~\cite{Minahan:2009te} for the $SL(2|1)$ multiplets and our SYT labels will be made more precise in section~\ref{section_SL(2|1)_sector}.}. Namely for each $m$ the unpaired $\cv^4_{2m,2}$ multiplet has the same energy as
the single
$\cv^2_{2m+1}$ multiplet,
and as we will see in section~\ref{short_cond} this is no accident. It is rather a manifestation of the fact that the two multiplets are short in the classical theory but they combine into a long multiplet in the interacting theory and so must have the same anomalous dimension.

In fact since (\ref{sl(2|1)_for_L=1}) has the symmetry $u_i\leftrightarrow v_i$ but there exists only one $\cv^2_{2m+1}$ in the $\cv^1\times\ocv^1$ tensor product, the corresponding solutions must necessarily have $u_i=v_i$ and hence the aforementioned equations become equivalent to
\begin{equation}\label{sl2BAE}
\frac{u_i+i/2}{u_i-i/2}=-\prod_{k=1,k\ne i}^{m}\frac{u_i-u_k-i}{u_i-u_k+i}\,.
\end{equation}
These are the BAE for the twist-1 states in the $SL(2)$ sector, first written
in~\cite{Gromov:2008qe}, reflecting the fact that a subset of $\cv^2_{2m+1}$ states belong to this sector.
We will refer to such BAE as ``$SL(2)$-like'' (see section~\ref{section_sl(2)_like_sectors} for more details).
The equations~(\ref{sl2BAE}) were subsequently solved
in~\cite{Zwiebel:2009vb}
and~\cite{Beccaria:2009ny} with the Baxter polynomial technique, and in particular the roots $u_i$ for the unique solution of~(\ref{sl2BAE}) are the solutions of the polynomial equation
\begin{equation}\label{baxter}
Q_m(u)={}_2F_1(-m, iu + \frac{1}{2}; 1; 2) = 0\,.
\end{equation}
The energy of the solution is shown to be
\begin{equation}\label{V2m,2}
\Delta_2=4 \big(S_1(m)-S_{-1}(m)\big)\,,
\end{equation}
and this concludes the proof that the unpaired states with $v_i=u_i$ belonging to the $\cv^4_{2m,2}$ multiplets have the same eigenvalues with the twist-1 states\footnote{We could arrive at the same conclusions for $\cv^4_{2m,2}$ in a relatively simpler manner, had we started with the BAE related to the Dynkin diagram of Figure~\ref{superD}(a), our choice however will simplify the analysis of the following sections.}. It is also worth mentioning that once we derive the $u$ roots from~(\ref{baxter}) we can replace them in~(\ref{rBAE}) (for $v_i=u_i$) to determine the $r$ roots and similarly obtain the $s$ roots from the $r$ roots through~(\ref{sBAE}).

\subsection{Proof of the $\cv^4_{4m+1,3}$ Eigenvalue Sequence}

Of course the results of the previous section could be straightforwardly obtained from multiplet splitting considerations, raising the question of whether the techniques we presented could be extended in order to find new eigenvalue sequences. The answer turns out to be positive, up to some extent analytically and further on numerically.

We first focus on the $\cv^4_{2m-1,3}$ multiplets, which apart from being the first obvious set to study since we just increase the second SYT label by one, they also have very similar root excitation structure,
\begin{equation}
[M_u,M_v,M_r,M_s,M_w]=[m,m,2m-2,2m-4,0]
\end{equation}
for $m\ge2$, namely they have just one fewer $s$ excitation as compared to the $\cv^4_{2m,2}$ multiplets. The Bethe equations for the $s$ roots will thus be identical to~(\ref{sBAE}), but now the problem is that they again correspond to a degree a $2m-3$ polynomial equation for the $s_i$, whereas there exist only $2m-4$ of them. We can get around this problem by restricting to unpaired states with $r_{M_r+1-i}=-r_i$ (and similarly for the $u, v$ and $s$ roots as well), in which case we see that~(\ref{sBAE}) is satisfied for $s_i=0$, the additional solution which does not correspond to an $s$ root. So similarly as before we can prove that
\begin{equation}
\prod_{k=1}^{2m-4}\frac{r_i-s_k-i/2}{r_i-s_k+i/2}=\frac{r_i+i/2}{r_i-i/2}\prod_{k=1,k\ne i}^{2m-2}\frac{r_i-r_k-i}{r_i-r_k+i}\,,
\end{equation}
which in turn simplifies the BAE for the $r$ roots  to
\begin{equation}\label{rBAE2}
\frac{r_i+i/2}{r_i-i/2}\prod_{k=1}^{m}\frac{r_i-u_k-i/2}{r_i-u_k+i/2}\prod_{k=1}^{m}\frac{r_i-v_k-i/2}{r_i-v_k+i/2}=1\,.
\end{equation}
For general $m$ we run into the same problem, as the above polynomial equation for $r_i$ is of degree $2m$ whereas there only exist $2m-2$ $r$ roots, namely there will be another two dual roots whose value we don't know a priori.

If however we focus on $m$ odd, then given that we are looking at unpaired states with $u_{M_r+1-i}=-u_i$, then the leftmost phase in~(\ref{rBAE2}) cancels that term in the first product corresponding to $u_i=0$. Now the polynomial equation becomes of degree $2m-2$ as it should, and once again using the theorem of appendix~\ref{useful_theorem} we obtain
\begin{equation}
\prod_{k=1}^{2m-2}\frac{u_i-r_k-i/2}{u_i-r_k+i/2}=\frac{u_i+i}{u_i-i}\prod_{k=1,k\ne i}^{m}\frac{u_i-u_k-i}{u_i-u_k+i}\prod_{k=1}^{m}\frac{u_i-v_k-i}{u_i-v_k+i}\,,
\end{equation}
and similarly for $v_j\leftrightarrow u_j$, from which the BAE for the $u$ and $v$ roots follow,
\begin{equation}
\begin{aligned}
\left(\frac{u_i+i/2}{u_i-i/2}\right)^2=\prod_{k=1}^{m-1}\frac{u_i-v_k-i}{u_i-v_k+i}\,,\\
\left(\frac{v_i+i/2}{v_i-i/2}\right)^2=\prod_{k=1}^{m-1}\frac{v_i-u_k-i}{v_i-u_k+i}\,.
\end{aligned}
\end{equation}
Here the product of $m-1$ terms implies that out of the $m$ roots we have excluded the root $u_i=0$, and since we have restricted ourselves to $m$ odd, consequently $m-1$ will always be even.

Finally if we consider the even more symmetric configuration of roots with $v_j=u_j$, the equations reduce again to the $SL(2)$-like
form~\cite{Gromov:2008qe} (this time for $L=2$)
\begin{align}
\left(\frac{u_i+i/2}{u_i-i/2}\right)^2=-\prod_{k=1,k\ne i}^{m-1}\frac{u_i-u_k-i}{u_i-u_k+i}\,,
\end{align}
the solution of which was also given
in~\cite{Beccaria:2009ny}. In this sector the two terms in the energy~(\ref{energies}) become equal so in order to determine the anomalous dimensions of the multiplets in question we just need to add 8, coming from the additional $u_i=0$ root, to the energy eigenvalues
of~\cite{Beccaria:2009ny}. Therefore we have proven that
\begin{equation}\label{energy_V4m+1,1}
\Delta_2=4S_1(m)+8\qquad\mbox{for $\cv^4_{4m+1,3}$}.
\end{equation}

\subsection{Numerical Methods for Other Eigenvalues}

As we saw in the previous section, $\cv^4_{2m-1,3}$ multiplets have just one fewer $s$ excitation than the $\cv^4_{2m,2}$ multiplets, which implies that their Bethe equations will be similar.
{}From that perspective, our observation that the $\cv^4_{2m,2}$ Bethe roots are quite close to satisfying the $\cv^4_{2m-1,3}$ BAE comes as no surprise, and it is thus natural to look for numerical solutions of the latter equations in this vicinity.

In fact, there is a whole series of multiplets with very similar excitation numbers. With the help of~(\ref{yt2dynkin}) and~(\ref{excitations_to_cartan}), it is easy to prove that a generic length-4 multiplet $\cv^4_{k_1,k_2}$ with $k_1\ge k_2\ge3$ has excitation numbers
\begin{equation} [M_u,M_v,M_r,M_s,M_w]=[\frac{k_1+k_2}{2}-1,\frac{k_1+k_2}{2}-1,k_1+k_2-4,k_1+k_2-6,k_2-3]\,,
\end{equation}
from which we conclude that all multiplets with the same $k_1+k_2$ will only differ in the number of $w$ roots. In particular starting with $M_w=0$ and increasing the number of $w$ roots we obtain the multiplets $\cv^4_{2m-1,3}, \cv^4_{2m-2,4}, \cv^4_{2m-3,5}, \ldots, \cv^4_{m+1,m+1}$.
So we can successively obtain the roots for all multiplets in this sequence by looking for numerical solutions at each step near the roots obtained in the previous step.

In more detail, we obtain the results of Table~\ref{tab4} in the following way:
\begin{enumerate}
\item We first calculate the $u$ roots for $\cv^4_{2m,2}$ from~(\ref{baxter}), and subsequently the corresponding $r$ and $s$ roots through~(\ref{rBAE}) (for $v_i=u_i$) and~(\ref{sBAE}).
\item We use Mathematica's \texttt{FindRoot[]} command to search for numerical solutions to the $\cv^4_{2m-1,3}$ and $\cv^4_{2m-2,4}$ BAE using the $\cv^4_{2m,2}$ Bethe roots determined in the previous step as starting points\footnote{Since we are looking at unpaired states, Dynkin roots with an odd number of excitations will necessarily have one zero Bethe root. Hence $\cv^4_{2m,2}, \cv^4_{2m-1,3}$ and $\cv^4_{2m-2,4}$ will all have the same number of unknown, positive Bethe roots.}. To avoid trouble when the search approaches singular values for the roots we have to express the BAE in polynomial rather than rational form.
\item The remaining multiplets in the set, $\cv^4_{2m-3,5}, \cv^4_{2m-4,6}$ and so on will have more $w$ roots and hence more unknowns. We estimate their starting values by plugging the values for the $u,r$ and $s$ roots of the previous step into the new BAE and solving for the $w_i$. In fact at each step we have to estimate only one $w$ root even though $M_w$ increases, because we notice that the remaining starting points can be very well estimated by the $w$ roots of the multiplets with $M_w$ smaller by 2, namely the ones calculated two steps back\footnote{In all cases we saw the $w_i$ are purely imaginary, so when selecting from all possible solutions for the estimated $w$ we can use this as a guiding principle. }.
\item Each set of multiplets whose roots we can determine in this iterative manner terminates with $\cv^4_{m+1,m+1}$, and so for a given $m$ we have filled a bottom-left to top-right diagonal line in Table~\ref{tab4}. If desired one could determine new sets of roots for higher values of $m$ indefinitely, and to arbitrarily high precision for the roots. The corresponding energy eigenvalues are then calculated via~(\ref{energies}).

\end{enumerate}

The identification of the rational sequences~(\ref{new_eigenvalue_sequences}) from the data generated by this algorithm is facilitated by first splitting each column into two pieces by taking the differences of next-to-consecutive eigenvalues in the same column, and then combining the two sets of results. For example in $\cv^4_{2m+1,3}$, where the answer was determined analytically for $m$ even~(\ref{energy_V4m+1,1}) and hinted at the existence of a simpler structure when $m$ increases by two, we separately identified the sequence $\Delta_2=4(S_1(m+1)+S_1(m))+8$ for $m$ odd and then moved on to combine the odd and even results. We should also note that the immense number of identities involving harmonic numbers allows us to rewrite (\ref{new_eigenvalue_sequences}) in many equivalent forms, for example we can replace the $1\pm(-1)^m$ terms in favor of more harmonic numbers, and/or combine formulas for more than one sequence to more general ones. We mention here the intriguing formula
\begin{equation}
\Delta_2=2\big(S_1(m)+S_{-1}(m)+S_1(m+1)+S_{-1}(m+1)\big)+4 \big(S_1(2p)-S_{-1}(2p)\big)\,,
\end{equation}
which holds exactly for $\cv^4_{2m+1,2p+1}$ with $m$ even for $p=0$, $m$ odd for $p=2$ and $m$ both odd and even for $p=1$, and more surprisingly it even gives good approximations to the irrational values of the remaining multiplets.

Another observation we can make is that the eigenvalues of the unpaired $\cv^4_{j,j}$ multiplets are just 4 times the Hamiltonian density $D_{123}$ eigenvalues, which can be obtained with the Hamiltonian diagonalization techniques we discuss in section~\ref{OSp(4|2)}. Since all length-3 $OSp(6|4)$ multiplets have $OSp(4|2)$
submultiplets~\cite{Papathanasiou:2009en}, it is sufficient to diagonalize this subset of states. This connection, apart from providing another way for calculating the unpaired $\cv^4_{j,j}$ eigenvalues from the simpler $OSp(4|2)$ sector, raises the question whether something similar holds for other cases as well, for example for the paired eigenvalues. It would also be interesting to find a group-theoretic proof of this relation.

\section{Shortening Conditions and Multiplet Splitting}\label{short_cond}

As is well known, for the states transforming in a representation of any superconformal group to respect unitarity certain inequalities between the scaling dimension and the remaining Cartan charges have to hold. Multiplets which saturate these inequalities turn out to have a large number of states with zero norm which can be consistently removed, resulting in a multiplet with a smaller number of positive norm states compared to generic, long multiplets.

In this section we classify all length-2 and -4 $OSp(6|4)$ supermultiplets according to which (if any) unitarity bounds they saturate, or in other words according to the shortening conditions they obey. For 3-dimensional superconformal groups these were first derived
in~\cite{Minwalla:1997ka} (see
also~\cite{Bhattacharya:2008zy}) and were further refined recently
in~\cite{Dolan:2008vc}. We will be using the classification and notations laid out in Table 1 of the latter paper, where the R-symmetry group is described in terms of Gelfand-Zetlin instead of Dynkin labels\footnote{For the case at hand the $SO(6)$ Gelfand-Zetlin labels $r_i$ are related to the $SU(4)$ Dynkin labels $d_i$ by $(r_1,r_2,r_3)=(d_2+\frac{1}{2}(d_3+d_1),\frac{1}{2}(d_3+d_1),\frac{1}{2}(d_3-d_1))$.\label{footnote_GZ_labels}}. Here we mention for reference that all semi-short representations and conserved currents obey $\Delta=r_1+j+1$, their difference being that the former additionally obey $r_1\ne0$, whereas the latter have $r_1=0$. BPS and $\frac{1}{2}$BPS multiplets obey $\Delta=r_1$ while all of the rest are long.

\TABLE[t]{
$\begin{array}{|c|c|c|c|c|c|l|c|}\hline
\mbox{Irrep}&\Delta&j&r_1&r_2&r_3&\mbox{Type}&\mbox{Denoted}\\\hline
\cv^2,\ocv^2&1&0&1&1&\mp1&\frac{1}{2}\mbox{BPS}, \pm&(3,B,\mp)\\\hline
\cv^2_1&1&0&1&1&0&\mbox{BPS}, n=2&(3,B,2)\\\hline
\cv^2_2&1&0&1&0&0&\mbox{BPS}, n=1&(3,B,1)\\\hline
\cv^2_k&\frac{k-1}{2}&\frac{k-3}{2}&0&0&0&\mbox{Conserved current}&(3,\mbox{cons.})\\\hline
\cv^4, \ocv^4&2&0&2&2&\mp2&\frac{1}{2}\mbox{BPS}, \pm&(3,B,\mp)\\\hline
\cv^4_1, \ocv^4_1&2&0&2&2&\mp1&\mbox{BPS}, n=2&(3,B,2)\\\hline
\cv^4_{1,1}&2&0&2&2&0&\mbox{BPS}, n=2&(3,B,2)\\\hline
\cv^4_{2,1}&2&0&2&1&0&\mbox{BPS}, n=1&(3,B,1)\\\hline
\cv^4_{2,2}&2&0&2&0&0&\mbox{BPS}, n=1&(3,B,1)\\\hline
\cv^4_{k}, \ocv^4_{k}&\frac{k+1}{2}&\frac{k-3}{2}&1&1&\mp1&\mbox{Semi-Short 2 \& 3}&(3,A,\mp)  \\\hline
\cv^4_{k,1}&\frac{k+1}{2}&\frac{k-3}{2}&1&1&0&\mbox{Semi-Short 1}, n=2&(3,A,2)  \\\hline
\cv^4_{k,2}&\frac{k+1}{2}&\frac{k-3}{2}&1&0&0&\mbox{Semi-Short 1}, n=1 &(3,A,1) \\\hline
\cv^4_{k_1,k_2}&\frac{k_1+k_2-2}{2}&\frac{k_1-k_2}{2}&0&0&0&\mbox{Long} &(3,A,0) \\\hline
\end{array}$
\caption{The classification of 2- and 4-site supermultiplets of the ABJM theory according
to~\cite{Dolan:2008vc}. The first row refers to the SYT labeling of the representations, whereas the $r_i$ correspond to the Gelfand-Zetlin indices of $SO(6)$, (see footnote~\ref{footnote_GZ_labels} for more details).}\label{tab3}
}

The classification is given in Table~\ref{tab3}. We see that for length-2 there exist no long or even semi-short multiplets, where as for length-4, there exist no conserved currents. It is also interesting to note that all multiplets which have representatives in the $OSp(4|2)$ sector, namely the $\cv^4_{j,p}$ multiplets with $p\le2$, are short. Consequently the $OSp(4|2)$ Hamiltonian diagonalization discussed in section~\ref{OSp(4|2)} can be used to find eigenvalues for all short $OSp(6|4)$ multiplets.

We should note that so far we have considered whether the classical dimension $\Delta$ satisfies the shortening conditions, since in the spin chain picture it is the dilatation operator that induces the correction $\Delta_2$. When we also include the latter, multiplets which are classically short will no longer saturate the unitarity bounds and hence they have to necessarily combine into long multiplets of the interacting theory. This process, which can be equivalently seen as the decomposition or splitting of the long interacting multiplet into short classical multiplets has been studied for the case of $OSp(2N,4)$ superconformal groups
in~\cite{Dolan:2008vc} as well. We mention here a particular decomposition of interest, respecting the notations of the latter reference,
\begin{equation}
\chi^{(3,{\rm long})}_{(j+1;j;0,0,0)}
=\chi^{(3,{\rm cons.})}_{(j+1;j;0,0,0)}
+\chi^{(3,A,1)}_{(j+\frac{3}{2};j-\frac{1}{2};1,0,0)}\, ,
\end{equation}
where subscripts refer to the quantum numbers $(\Delta,j;r_1,r_2,r_3)$ and superscripts to the type of the multiplet as in rows 2-6 and row 8 of Table~\ref{tab3} respectively. Comparing with the SYT notation of the leftmost row we learn that the short multiplets $\cv^2_{2m+1}$ and $\cv^4_{2m,2}$ ($m=j-1$) combine to form a long multiplet, and for this joining to occur the anomalous dimensions of the two short multiplets must be equal. This explains the reduction of the $\cv^4_{2m,2}$ BAE to those of $\cv^2_{2m+1}$ that we saw in section~\ref{V2m}.

\section{Multiplets in Various Subsectors}

\subsection{$OSp(4|2)$ Sector}\label{OSp(4|2)}

The two-loop dilatation operator acting on a spin
chain state of length $2L$ in the ABJM theory in the R-matrix form is given
by~\cite{Zwiebel:2009vb,Minahan:2009te}
\begin{equation}
\label{eq:fulldtwo}
\Delta_2  = \sum_{i=1}^{2 L} (D_2)_{i,i+1,i+2}\,
\end{equation}
where the Hamiltonian density $D_2$ acts simultaneously on three adjacent
sites of the chain according to\footnote{Harmonic numbers with half-integer arguments are related to ordinary harmonic numbers by the identity $S_1(j-1/2)+2\log2=2S_1(2j)-S_1(j)$, which also implies that all coefficients in the second sum of~(\ref{dtwo}) will be rational.}
\begin{multline}
(D_2)_{123} = \sum_{j=0}^\infty S_1(j)
{\cal P}^{(j)}_{12} \\
+ \sum_{j_1,j_2,j_3 = 0}^\infty
(-1)^{j_1+j_3} \left({\textstyle{\frac{1}{2}}} S_1(j_2-{\textstyle{\frac{1}{2}}})
+ \log 2 \right)
\left(
{\cal P}^{(j_1)}_{12}
{\cal P}^{(j_2-1/2)}_{13}
{\cal P}_{12}^{(j_3)} +
{\cal P}^{(j_1)}_{23}
{\cal P}^{(j_2-1/2)}_{13}
{\cal P}^{(j_3)}_{23}
\right).
\label{dtwo}
\end{multline}
Here ${\cal P}_{ab}^{(j)}$ is the projection operator that acts on sites $a$ and $b$ and equals the identity if the length-2 state belongs to an irreducible multiplet labeled by $OSp(6|4)$ spin $j$, and zero otherwise. In more detail, $j$ is related to the eigenvalue of the $OSp(6|4)$ Casimir operator $J^2=j(j+1)$, expressed in terms of the Cartan labels of the multiplet in question as
\begin{equation*}
J^2={\textstyle \frac{1}{2}} \Big(\Delta (\Delta+3)+j(j+1)-{\textstyle\frac{1}{4}}d_1(d_1+2)-{\textstyle\frac{1}{4}}d_3(d_3+2)-{\textstyle\frac{1}{8}}(2d_2+d_1+d_3)^2-(2d_2+d_1+d_3))\Big)\,,
\end{equation*}
which through~(\ref{yt2dynkin}) can be expressed in terms of the SYT labels as well. For the relevant length-2 multiplets $\cv^2_k$
(see~\cite{Papathanasiou:2009en} for more details) $k$ and $j$ turn out to be related simply by $j=2k+1$, which also demonstrates why $j$ uniquely characterizes a multiplet appearing in the $\cv^1\times\ocv^1$ or $\cv^1\times\cv^1$ decompositions.

An explicit form of the projectors ${\cal P}_{ab}^{(j)}$ has not been worked out for the full $OSp(6|4)$ group, but it has been derived for the $OSp(4|2)$
sector~\cite{Zwiebel:2009vb} by looking at states containing fields from the following subsets of the two singleton representations,
\begin{equation}\label{osp42fields}
\begin{aligned}
\phi_i\;, \bar\psi^{4-i+1},\quad i=1,2&&\mbox{for } \cv^1,\\
\bar\phi^{4-i+1}, \psi_{i}\;,\quad i=1,2&&\mbox{for } \ocv^1
\end{aligned}
\end{equation}
where we only keep the first Lorentz component of the fermions, and just one covariant derivative ${\cal D}_{11}$ in the corresponding direction, acting on both fields.

In this sector~(\ref{dtwo}) continues to hold, and so we can obtain the anomalous dimensions of $OSp(4|2)$ multiplets by explicit diagonalization of the dilatation operator. Our case of interest will be the shortest states the dilatation operator in the form~(\ref{dtwo}) can act on, namely of length 4. The diagonalization can be performed in any set of states which are closed under the action of~(\ref{dtwo}), and for our purposes it will suffice to consider all states with a given classical scaling dimension $\Delta$.

Another operator that can easily be seen to commute with the Hamiltonian $\Delta_2$ (and also the $OSp(6|4)$ Cartan charges) is the translation operator $T$ that sends site $i$ to $i+1$,
\begin{equation}
T|A_1 B_1 A_2\cdots A_L B_L\rangle=
(-1)^{\deg(A_1) \deg(B_1 A_2 B_2 \cdots A_L B_L)}|B_1 A_2 B_2 \cdots A_L B_L A_1\rangle \,.
\end{equation}
However since the action of $T$ is from a $(\cv\times\ocv)^L$ Fock space to a $(\ocv\times\cv)^L$ one, its square $T^2$ that shifts each site by two will be more relevant. We can thus find a common basis of eigenstates, and for length-4 states the possible eigenvalues of $T^2$ will be $\pm1$ since $T^4=1$. In fact, we can diagonalize the combination $T^2 \Delta_2$ to obtain the $T^2$ and $\Delta_2$ the eigenvalues simultaneously since the latter are always positive.

Our results are summarized in Table~\ref{tab1}, where the $\Delta_2$ eigenvalues have been distributed to the different multiplets, and we have included both $T^2=\pm1$ cases for completeness, even though physical, cyclically invariant states have $T^2=1$ only\footnote{As a consistency check, we notice that for each multiplet, the total number of eigenvalues is equal to its multiplicity in the tensor product decompositions~(\ref{f=4decomp1}),~(\ref{f=4sym_decomp}). }. Since a $OSp(4|2)$ representation will belong to a larger representation of the full superconformal group, states which are primary in $OSp(4|2)$ will generically be descendants in $OSp(6|4)$. This is why we have chosen to label the representations in terms of the charges of the corresponding $OSp(6|4)$ primary.
{}From those we can obtain the charges of the $OSp(4|2)$ primary state in the following manner; for $\Delta$ (or equivalently $j$), we have to shift all $\cv^4_{2m+1,1}, \cv^4_{2m+2}, \cv^4_{2m+2,2}$ for $m\ge1$ by $\frac{1}{2}$, and leave the remaining multiplets the same, whereas the $SO(4)$ charges are obtained from the $SU(4)$ charges simply by dropping the middle label.

We have used numbers with two decimal digits to denote irrational eigenvalues, which we see appear already at $\Delta=3.5$ ($\Delta=4$ if one restricts to cyclic states).
All irrational eigenvalues correspond to roots of polynomial equations, and generically one cannot express them in terms of a nice closed form. For example the three eigenvalues of $\cv^4_8$
are solutions of the cubic polynomial equation $x^3-32x^2+\frac{1669}{5}x-\frac{152894}{135}$.

\begin{landscape}
\TABLE{
$\begin{array}{|c|c|c|l|l|}\hline
\Delta&SU(4)&\mbox{SYT}&\Delta_2, T^2=+1&\Delta_2,T^2=-1\\\hline
2&[2,0,2]&\cv^4_{1,1}&0& \\
 &[2,0,0]&\cv^4_2&&4 \hfill+\mathord{\mbox{conj.}}\\
&[0,2,0]&\cv^4_{2,2}&8& \\
&[1,0,1]&\cv^4_{3,1}&6^2&4^2  \\
\hline
\frac{5}{2}&[2,0,0]&\cv^4_4&6^2& 28/3 \hfill+\mathord{\mbox{conj.}}\\
 &[0,1,0]&\cv^4_{4,2}&8&(22/3)^2, 10^2\\
 \hline
3&[1,0,1]&\cv^4_{5,1}&4, 8^2, 11^2&(22/3)^4 \\
\hline
\frac{7}{2}&[2,0,0]&\cv^4_6&10^2&6.61, (25/3)^2, 12.86 \hfill+\mathord{\mbox{conj.}}\\
  &[0,1,0]&\cv^4_{6,2}&(25/3)^4, 32/3, 12^2&10^2, (182/15)^2  \\
  \hline
4&[1,0,1]&\cv^4_{7,1}&(7.94)^2, (134/15)^4,(14.06)^2&(6.26)^2, (173/15)^4, (11.07)^2  \\
\hline
\frac{9}{2}&[2,0,0]&\cv^4_8&(7.87)^2,(10.76)^2, (13.36)^2 &(157/15)^2, 10.87,15.30 \hfill+\mathord{\mbox{conj.}}\\
  &[0,1,0]&\cv^4_{8,2}&32/3, 12^2, (63/5)^4&(8.89)^2, (143/15)^4, (11.88)^2, (13.79)^2,(14.95)^2 \\
  \hline
5&[1,0,1]&\cv^4_{9,1}&6, (28/3)^2, (247/21)^4, (12.28)^2, (40/3)^2, (16.22)^2&(8.87)^4, (10.82)^4, (14.52)^4  \\
\hline
\frac{11}{2} &[2,0,0]&\cv^4_{10}&(11.25)^2,(79/7)^2, (15.75)^2 &8.13, (9.69)^2, (12.96)^2, (13.52)^2, 14.01, 17.17 \hfill+\mathord{\mbox{conj.}}\\
  &[0,1,0]&\cv^4_{10,2}&(9.62)^4, (11.03)^4, 184/15, (40/3)^2, (46/3)^2, 15.38^4&(11.88)^2, (447/35)^4, (13.23)^2, (13.79)^2, (16.96)^2 \\
  \hline
\end{array}$
\caption{Two-loop planar anomalous dimensions $\Delta_2$ of low-lying states of length-4 in the $OSp(4|2)$ sector of the ABJM theory. The Cartan labels correspond to the charges of the
$OSp(6|4)$, rather than the $OSp(4|2)$, primary state of the multiplet, and the $SO(3)$ charge is always given by $j=\Delta-2$ in this sector. The SYT column refers to the super-Young
tableau labeling of the multiplets. Superscripts denote the multiplicities of eigenvalues, when they are larger than one.
The label `+conj.' refers to the entire line and represents conjugate states with
$SU(4)$ labels reversed.}\label{tab1}
}
\end{landscape}

The results of Table~\ref{tab1} were obtained by explicit
diagonalization of~(\ref{eq:fulldtwo}) within the (finite-dimensional)
subspace of states for various $\Delta$.
At each level $\Delta$, the $\Delta_2$ eigenvalues will come both from descendant and primary states. The eigenvalues for the descendant states are known from their primaries that have appeared at a previous step with lower $\Delta$, and so we can remove as many eigenvalues as the number of descendants we expect at this level. The number of descendants can in turn be found with the help of the 4-fold tensor product decompositions~(\ref{f=4decomp1}),~(\ref{f=4sym_decomp}) (restricted to $OSp(4|2)$) and the relevant characters for the length-4 $OSp(4|2)$ multiplets
from~\cite{Papathanasiou:2009en}. We attribute the remaining eigenvalues to new multiplets once more by comparing their multiplicities with the number of primary states that are expected from the decomposition and character formulas. This iterative process we have described is a particular realization of the so-called ``Eratosthenes' supersieve'' technique
(see~\cite{Bianchi:2003wx,Beisert:2003te,Beisert:2004ry}).

We should mention that we have not yet exhausted the set of charges which commute with the Hamiltonian. Another such charge is the spin chain parity
$\boldsymbol{p}$~\cite{Zwiebel:2009vb}, which acts on a state by reversing the order of the spin chain sites (with a minus sign if an odd number of fermions cross each other),
\begin{equation}
\boldsymbol{p}|A_1 B_1 A_2\cdots A_L B_L\rangle=
(-1)^{n_f(n_f-1)/2}|B_L A_L \cdots B_1 A_1\rangle \,,
\end{equation}
where $n_f$ is the total number of fermions in the state. For the length-4 states we focus on it is easy to show that the spin chain parity commutes with $T^2$ as well, and more generally it commutes with the operator $\frac{1}{L} \sum_{a=1}^L T^{2a}$ that projects to cyclically invariant states for any length. Even though acting with $\boldsymbol{p}$ changes the type of representation at each site, we can combine it with $T$ to get an operator that maps states within the same $(\cv^1\times\ocv^1)^L$ Fock space. This operator $\boldsymbol{p_T}\equiv\boldsymbol{p}T$, which we could perhaps call ``shifted parity'', inherits the commutation relations of $\boldsymbol{p}$, and obeys $\boldsymbol{p_T}^2=1$. Hence it can be used to label states with an additional (+) or (-) sign, similar to the plain parity operator in ${\cal N}=4$ SYM.

An important observation related to our discussion of parity is that the multiplicity of the eigenvalues is either even or 1, no other odd values appear. Namely almost the entire spectrum is arranged in pairs of states with degenerate energies, with the only exception coming from unpaired states. With our definition of $\boldsymbol{p_T}$ we can additionally check that all pairs have opposite parities, and it is the combination of parity symmetry and integrability that accounts for this phenomenon. In particular, the third integrable charge $Q_3$ commutes with the Hamiltonian $\Delta_2$ and anticommutes with $\boldsymbol{p_T}$, so that acting on a common eigenstate of the three operators with $Q_3$ creates a state with opposite parity (from the anticommutation with $\boldsymbol{p_T}$) and the same energy (from the commutation with $\Delta_2$). The only exception to this pairing is when the initial state is annihilated by $Q_3$, which accounts for the existence of unpaired states in the spectrum.
See~\cite{Beisert:2004ry} for a discussion of parity in the context of ${\cal N}=4$ SYM.

Focusing on the unpaired states of the $OSp(4|2)$ sector, we notice that in the cyclically invariant part we find one for each level of $m$ in the $\cv^4_{4m+1,1}$ and $\cv^4_{2m,2}$ multiplets only\footnote{For completeness, we mention that the unpaired noncyclic $\cv^4_{2m}$ states and their conjugates $\ocv^4_{2m}$ have shifted parities (-) and (+) respectively.}, with their parities being (+) and (-) respectively, which perhaps suggests an alternating pattern as we increase the second SYT label. As an additional check, we verify that their eigenvalues \{0,4,6,\ldots\} and \{8,8,32/3,32/3,184/15\ldots\} indeed agree with the sequences~(\ref{V4m+1,1}) and~(\ref{V2m,2}) respectively.

\subsection{$SL(2|1)$ Sector}\label{section_SL(2|1)_sector}

The $SL(2|1)\simeq OSp(2|2)$
sector~\cite{Minahan:2009te} contains fields at each site which are obtained by (\ref{osp42fields}) for $i=1$, and is thus a subsector of the $OSp(4|2)$ sector. In the notation of the latter paper, the corresponding algebra apart from the usual $SL(2)$ generators
\begin{equation}\label{SL(2|1)_algebra1}
[J_0,J_\pm]=\pm J_\pm\qquad[J_+,J_-]=2 J_0,
\end{equation}
also has a second antihermitian Cartan charge $H$ and four fermionic charges, $Q^+,Q^-, S^+=-(Q^-)^\dagger,S^-=-(Q^+)^\dagger$ with (anti)commutation relations
\begin{align}\label{SL(2|1)_algebra2}
[J_0,Q^\pm]&=\frac{1}{2}Q^\pm & \{Q^+,Q^-\}&=J_+&[J_-,Q^\pm]&=S^\pm\nonumber\\
[H_0,Q^\pm]&=\pm\frac{1}{2}Q^\pm & \{Q^+,S^-\}&=H-J_0,&&
\end{align}
together with what can be obtained from the above by hermitian conjugation. It is trivial to show how the $SL(2|1)$ algebra in this basis is embedded into the full $OSp(6|4)$ algebra of appendix~\ref{osp64_appendix},
\begin{align}
J_0&=\frac{1}{2}I^1_{\phantom{1}1}&J^+&=\frac{1}{2}P^{11}&Q^+&=\frac{1}{\sqrt{2}}S^{11}&S^+&=-\frac{1}{\sqrt{2}}M_1^{\phantom{1}1}\nonumber\\
H&=\frac{1}{2}U^1_{\phantom{1}1}&J^-&=-\frac{1}{2}K_{11}&Q^-&=\frac{1}{\sqrt{2}}M^1_{\phantom{1}1}&S^-&=-\frac{1}{\sqrt{2}}S_{11},
\end{align}
from which we see that the $SL(2|1)$ sector includes all states that can be constructed from superoscillators whose bosonic and fermionic indices each take only a single value.

{}From this point on we can use the oscillator method \cite {Gunaydin:1988kz} (see also \cite{Gunaydin:1984wc,Gunaydin:1985tc,Gunaydin:1998jc,Fernando:2004jt} for other applications in supergravity and ${\cal N}=4$ SYM) to determine what subset of $OSp(6|4)$ representations appears in this sector and relate their $(j,h)$ charges, corresponding to the $(J_0,H)$ generators, with the alternative labels we have been using throughout this paper. Finding these representations is equivalent to investigating how a particular graded (anti)symmetrized combination of superoscillators, corresponding to the lowest weight of the representation and encoded in its SYT, can decompose into bosonic and fermionic oscillators. For $SL(2|1)$ this is particularly simple since restricting to an oscillator whose index can only have one value means that we can only have symmetric combinations. Then, translating our results to any system of Cartan labels requires expressing the corresponding charges in terms of number operators and reading their value for the primary state of the multiplet
(see~\cite{Papathanasiou:2009en} for additional information).

In this manner we find that for any length $f$ the $SL(2|1)$ multiplets that appear are, in terms of both $(j,h)$ and the usual embedding $OSp(6|4)$ charges,
\bigskip
\begin{center}
\begin{tabularx}{\textwidth}{
>{\hsize=0.3\hsize\centering\arraybackslash}X
|>{\hsize=0.5\hsize\centering\arraybackslash}X
|>{\hsize=1.05\hsize\centering\arraybackslash}X}
SYT&$(2j,2h)$&$[\Delta,j,d_1,d_2,d_3]$\\\hline
$\cv^f$&$(\frac{f}{2},-\frac{f}{2})$&$[\frac{f}{2},0,f,0,0]$ \\
$\cv^f_{k,\underbrace{1,\ldots,1}_m}$&$(k-1+\frac{f}{2},m+1-\frac{f}{2})$&$[\frac{1}{2}(k+1-f),\frac{1}{2}(k-1),f-m-1,0,m+1]$
\end{tabularx}
\end{center}
\bigskip
\noindent
where $m\le f-1$. We notice that for each multiplet one can take its conjugate with $d_1\leftrightarrow d_3$ by replacing $m\to f-m-2$, which in the $(j,h)$ charges translates into flipping the sign of $h$. The Cartan labels on the right hand side now refer to $SL(2|1)$, not $OSp(6|4)$,  primaries. Finally the existence of only a subset of $OSp(6|4)$ multiplets in the $SL(2|1)$ sector also simplifies the four-fold (symmetric) tensor product decompositions, which can be obtained from~(\ref{f=4decomp1}) and~(\ref{f=4sym_decomp}) by simply dropping all multiplets whose second SYT label is greater than or equal to 2.

A nice feature of the $SL(2|1)$ sector is that its Bethe equations (in a certain Dynkin basis) become much simpler (as does, of course, the direct Hamiltonian
diagonalization).
In order to see this, one can start with the BAE corresponding to the distinguished Dynkin diagram, (\ref{betheeq}), dualizing the fermionic root $s$ according to the method explained in appendix~\ref{useful_theorem} and then dualizing the $r$ root (which becomes fermionic in the first dualization). In this way we arrive at a symmetric Cartan matrix of the form
\begin{equation}
{\cal{K}} = \left(\begin{array}{ccc|cc}
& -1 &&&\\
-1&+2 &-1&&\\
& -1 &&+1&+1\\\hline
&&+1&&-2\\
&&+1&-2&
\end{array}\right),
\label{Cartan_matrix}
\end{equation}
where the columns from left to right correspond to the $w,\tilde s,\tilde r,v, u$ roots (tildes denote the dual roots), encoded in the Dynkin diagram of in Figure~\ref{superD}(a). It is straightforward to show that the $SL(2|1)$ algebra corresponds to the $2\times2$ lower right corner of the Cartan matrix, with $S^\pm$ the lowering, $Q^\pm$ the raising and $2(\pm H_0-J_0)$ the Cartan generators respectively.

\FIGURE[t]{
\centerline{\includegraphics[width=10cm]{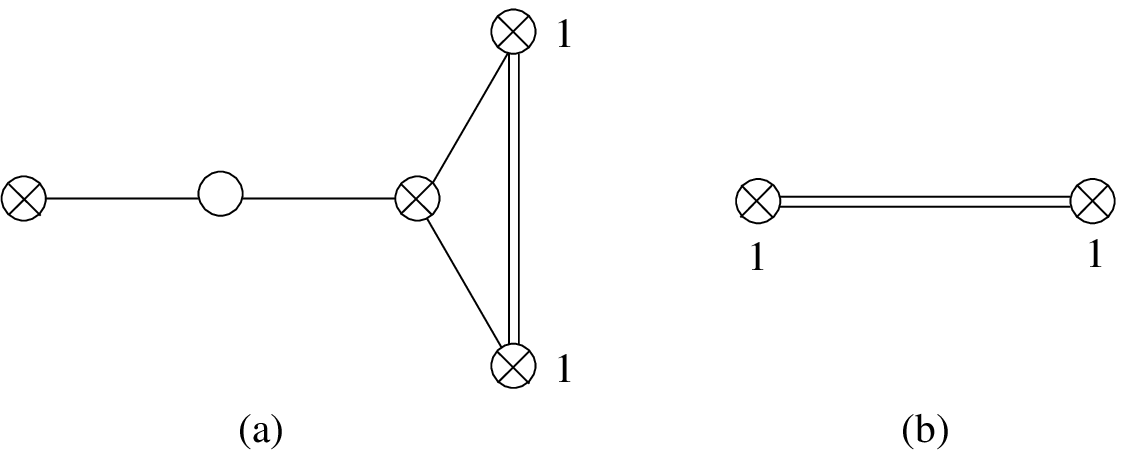}}
\caption{\label{superD}\small Super-Dynkin diagrams for
(a) $OSp(6|4)$ (b) $SL(2|1)$. }
}

Hence the $SL(2|1)$ Dynkin diagram reduces to that of Figure~\ref{superD}(a), giving rise to the simple BAE
\begin{equation}\label{sl(2|1)_BAE}
\begin{aligned}
\left(\frac{u_i+i/2}{u_i-i/2}\right)^L=\prod_{k=1}^{M_v}\frac{u_i-v_k-i}{u_i-v_k+i}\,,\\
\left(\frac{v_i+i/2}{v_i-i/2}\right)^L=\prod_{k=1}^{M_u}\frac{v_i-u_k-i}{v_i-u_k+i}\,,
\end{aligned}
\end{equation}
which we already encountered in section~\ref{chapter_unpaired_multiplets} for the $f=2L=4$ case of interest. Since we would like to know which multiplet it is whose roots are to  be calculated with~(\ref{sl(2|1)_BAE}), the next step is to determine the relation between the excitation numbers $[M_u,M_v]$ and the Cartan charges $(j,h)$. In the oscillator notation of appendix~\ref{osp64_appendix}
and~\cite{Papathanasiou:2009en}, the $SL(2|1)$ Cartan generators are expressed in terms of number operators as $(J_0,H)=(N_{B_1}/2+f/4,N_{F_1}/2-f/4)$, whereas the ground state $\mbox{Tr}\big[(\phi_1\bar\phi^4)^L\big]$ is equivalent to the tensor product $(|0\rangle\times \gamma^1|0\rangle)^L$, and hence has charges $(j,h)=(f/4,f/4)$.
{}From this we infer that a state with excitation numbers $[M_u,M_v]$ has
charges
\begin{equation}
(j,h)=(\frac{1}{2}(M_u+M_v+L),\frac{1}{2}(M_u-M_v))\,.
\end{equation}
In terms of the SYT notation this means that the $\cv^f$ multiplets do not appear as excitations of the aforementioned ground state, whereas the remaining multiplets have
\begin{equation}\label{sl21_syt2excitations}
\cv^f_{k,\underbrace{1,\ldots,1}_m} : k=1+M_u+M_v,\quad m=M_u-M_v+L-1.
\end{equation}

\subsection{$SL(2)$ (-like) Sectors}\label{section_sl(2)_like_sectors}

It is worth mentioning that there exist two smaller subsectors of the $SL(2|1)$ sector which are closed under the action of
$\Delta_2$~\cite{Zwiebel:2009vb}, even though their ground states are not the primary states $\mbox{Tr}\big[(\phi_1\bar\phi^4)^L\big]$.

In particular these are the two $SL(2)$ subsectors with ground state the descendant $\mbox{Tr}\big[(\phi_1 \psi_1)^L\big]$ (resp. $\mbox{Tr}\big[(\bar\psi^4 \bar\phi^4)^L\big]$), and excited states having an arbitrary number of $n$ symmetrized derivatives ${\cal D}_{11}$ that belong to the multiplets $\cv^{2L}_{L+2n}$  (resp. $\ocv^{2L}_{L+2n}$), and excitation numbers  $[n,n+L-1]$ (resp. $[n+L-1,n]$) due to (\ref{sl21_syt2excitations}).

Another interesting subset of states in the $SL(2|1)$ sector are those with $M_u=M_v=M$ and $u_i=v_i$, which as we have mentioned are described by the $SL(2)$-like\footnote{Namely, it is identical to the $SL(2)$, or $SU(1,1)$, sector of ${\cal{N}}=4$ SYM except for the overall minus sign on the right-hand side.} BAE
\begin{align}\label{GV-sl2}
\left(\frac{u_i+i/2}{u_i-i/2}\right)^L=-\prod_{k=1,k\ne i}^{M}\frac{u_i-u_k-i}{u_i-u_k+i}\,.
\end{align}
These states necessarily have $(j,h)=(M+L/2,0)$, and hence in SYT notation they correspond to the multiplets $\cv^{2L}_{k,1,\ldots,1}$ with $m=L-1$ labels equal to $1$ and $k=2M+1$.  We notice that states with this simplified BAE belong to the abovementioned $SL(2)$ subsectors only for $L=1$. It is also straightforward to check that the form of the projectors in~(\ref{dtwo}) is such that the dilatation operator mixes states with different numbers of fermions for any other ordering of the $SL(2|1)$ fields.
Hence there does not exist any other $SL(2)$ sector and so strictly
speaking the simplified BAE~(\ref{GV-sl2}) describes a particular set of states
rather than an $SL(2)$ sector.

A final observation for the BAE~(\ref{GV-sl2}) is that the total number $N_{\rm tot}$ of regular solutions (not restricted to cyclic ones) seems to depend on $L$ and $M$ as
\begin{equation}\label{sl(2)_solution_number}
N_{\rm tot}=\frac{(L-1+M)!}{(L-1)!M!}=\left(
\begin{array}{c}
L-1+M\\
M
\end{array}
\right).
\end{equation}
Comparing with (\ref{f=4decomp1}) we see that from a total of $j^2$ $\cv^4_{2j-1,1}$ multiplets, $j$ of them will be described by the $SL(2)$-like BAE (\ref{GV-sl2}), namely will have $u_i=v_i$ (but are not necessarily unpaired), whereas the remaining $j(j-1)$ will have $u_i\ne v_i$.

\section{Some Comments on Length-6 Operators}

In this section we sketch a preliminary analysis of length-6 states. For simplicity we focus on the unpaired states of the $SL(2|1)$ sector, which from what we saw in section~\ref{section_sl(2)_like_sectors} correspond to the multiplets $\cv^6_{2M+1,1,1}$.

As for the case of length 4, a guiding principle for attributing the eigenvalues of the dilatation operator to certain multiplets is the 6-fold tensor product decomposition of singleton representations.  After presenting this result we discuss a pattern that
we have found in the sum of unpaired eigenvalues of like
representations (i.e., the trace of
the dilatation operator in a certain subspace).

\subsection{6-fold Tensor Product Decomposition in the $SL(2|1)$  Sector}

The proof of the 6-fold tensor product decomposition is similar to the 4-fold case that we investigated
in~\cite{Papathanasiou:2009en}, and it requires character formulas for $SL(2|1)\simeq OSp(2|2)$ representations, which we have obtained
from~\cite{Dolan:2008vc}. We find that
\begin{equation}
\begin{aligned}
(\cv^1\otimes \ocv^1)^3&=\sum_{j=0}^\infty \binom{j+4}{4}(\cv^6_{2j+3}+\ocv^6_{2j+3})\\
&\qquad+(j+2) \binom{j+3}{3}(\cv^6_{2j+2,1}+\ocv^6_{2j+2,1})\\
&\qquad+\binom{j+2}{2}^2\cv^6_{2j+1,1,1}\,,
\end{aligned}
\end{equation}
where we can translate to different notations for the multiplets as explained in section~\ref{section_SL(2|1)_sector}.

Once again, comparison with (\ref{sl(2)_solution_number}), reveals that $(j+1)(j+2)/2$ of the $\cv^6_{2j+1,1,1}$ multiplets will have equal $u$ and $v$ roots, whereas the remaining $j(j+3)/2$ won't.
\TABLE{
$\begin{array}{|c|c|l|}\hline
M&D_2, \mbox{unpaired}&\mbox{Sum}\\\hline
0&0& 0 \\\hline
1&8& 8 \\\hline
2&2.34, 13.65& 16 \\
\hline
3&9.12, 17.54& 80/3 \\\hline
4&3.75, 14.44, 20.47& 116/3  \\
\hline
5&9.92, 18.20, 22.82& 764/15 \\\hline
6&4.76, 15.04, 21.02, 24.78& 328/5\\\hline
7&10.55, 18.71, 23.29, 26.46&8296/105\\\hline
\end{array}$
\caption{Two-loop planar anomalous dimensions $D_2$ for unpaired low-lying states of length 6, belonging to $\cv^6_{2M+1,1,1}$ multiplets.}\label{length-6}
}

\subsection{Sums of Unpaired Eigenvalues}

We have obtained the unpaired (multiplicity 1) eigenvalues for low-lying states of length-6 in the $SL(2|1)$ sector,first by Hamiltonian diagonalization, and as an independent cross-check by numerically solving the $SL(2)$-like BAE~(\ref{GV-sl2}). The first method is slightly more advantageous since we can simultaneously obtain the ``shifted parity'' $\boldsymbol{p}_T$ eigenvalue as well, which always turns to out to be $(+)$.

The results are shown in Table~\ref{length-6} where the integer $M$, apart from denoting the representations the unpaired states belong to ($\cv^6_{2M+1,1,1}$), is equal to the number of excitations in~(\ref{GV-sl2}). Irrational eigenvalues, which we approximate with two decimal digits, already appear at $M=2$, and we notice that the number of unpaired states now grows with $M$ as $[M/2+1]$. This renders the identification of eigenvalue sequences rather difficult. However the sum of all unpaired eigenvalues at each $M$ remains rational.

Both the number of solutions and the values for the energies suggest that their sum for each $M$, which we'll call $q(M)$, belongs to two different rational sequences, for odd and even $M$. Indeed we find that our results are consistent with the sequences
\begin{equation}
\begin{aligned}
q(M)&=4(M+1)S_1(M)+2(M+3) S_1(\frac{M}{2})-6 M&\mbox{if $M$ even},\\
q(M)&=4(M+2)S_1(M+1)+2 MS_1(\frac{M+1}{2})-6 (M+1)&\mbox{if $M$ odd}.
\end{aligned}
\end{equation}
As it was the case for paired length-4 states, the length-6 unpaired eigenvalues can be obtained by a polynomial equation, this time of degree $[M/2+1]$, for which $q(M)$ must therefore be the constant term. Perhaps the identification of the remaining coefficients is also possible, though challenging as they increase with $M$ much faster than $q(M)$.

\section{Outlook and Open Questions}

In this paper we have studied the two-loop spectroscopy of (primarily length-4) operators in planar ABJM theory. We were able to calculate the anomalous dimensions of all operators with classical dimension $\Delta\le 11/2$ in the $OSp(4|2)$ sector, and, more generally developed a method that allows us to determine the Bethe roots of one unpaired state for each $OSp(6|4)$ representation that appears in the Fock space of length-4 operators.
{}From the data obtained in this manner we identified three new sequences of rational eigenvalues.

In the analysis of section~\ref{chapter_unpaired_multiplets} we only looked for regular unpaired solutions to the Bethe ansatz, so it would be interesting to also investigate singular ones. To that end it would be useful to determine the number of unpaired solutions for a given set of excitation numbers. For example, from the tensor product decomposition (\ref{f=4sym_decomp}) we see that the $\cv^4_{2j,4p}$ and $\cv^4_{4j-1,2p-1}$ multiplets always appear an even number of times, so at least one more of each one must be unpaired.  We note that in the particular case of $\cv^4_{3,3}$ it was
indeed found~\cite{Minahan:2008hf} that the other unpaired solution is
singular.

It would also be nice to extend our analysis to higher loops. In particular we could derive the next-to-leading (NLO) order Baxter polynomial for the $\cv^2_{2m+1,1}$ (equivalently, unpaired $\cv^4_{2m,2}$) multiplets according to the methods
of~\cite{Kotikov:2008pv}, or even use the known NLO Baxter polynomial
of~\cite{Beccaria:2009ny} for the unpaired $\cv^4_{2m+1,1}$ multiplets, and search for numerical solutions of the Bethe equations corresponding to other multiplets, in the vicinity of its roots. At this order wrapping effects start to appear, which we would have to address according to the
proposal~\cite{Gromov:2009tv}.

\acknowledgments{
We thank M.~Beccaria and B.~Zwiebel for helpful correspondence
in the initial stages of this work.
G.~P. is grateful to the organizers of the Mathematica Workshop at the
University of Porto, Strings 2009 in Rome and the Integrability in Gauge
and String Theory
Workshop at the AEI Potsdam for hospitality and support
during the course of this work.
This work was supported in part by the
Department of Energy under contract DE-FG02-91ER40688 Task J and
the National Science Foundation under grant
PHY-0638520.
}

\appendix

\section{Fermionic Root Dualization}\label{useful_theorem}

\subsection{A Useful Example}
Let us start by reviewing a simple argument presented
in~\cite{Staudacher:2004tk} (also independently discovered by K.~Zarembo) which plays an important
role in the analysis of section~\ref{chapter_unpaired_multiplets}.

\bigskip
\noindent
{\bf Statement:} The equation
\begin{equation}\label{simpleBAE}
\prod_{k=1}^n\frac{x-y_k+a}{x-y_k-a}=1
\end{equation}
for the variable $x$
implies the relation
\begin{equation}\label{nice_relation}
\prod_{k=1}^{n-1}(y_i-x_k+a)=\frac{1}{2a n}\prod_{k=1}^{n-1}(y_i-y_k+2a)=\frac{1}{ n}\prod_{k=1,k\ne i}^{n-1}(y_i-y_k+2a)\,,
\end{equation}
where $x_i,\, i=1,2,...,n-1$ are the solutions of~(\ref{simpleBAE}).

\bigskip
\noindent
{\bf Proof:} Consider the polynomial
\begin{equation}
P(x)=\prod_{k=1}^{n}(x-y_k+a)-\prod_{k=1}^{n}(x-y_k-a)\,.
\end{equation}
It is of degree $n-1$ since the term $x^n$ cancels out between the two products, and the coefficient of the $x^{n-1}$ term can be easily evaluated with the use of Viet\'e's formulas
\begin{equation}
c_{n-1}=-(\sum_{k=1}^n y_k - n a- \sum_{k=1}^n y_k- n a)=2n a\,.
\end{equation}
Equation~(\ref{simpleBAE}) is equivalent to $P(x)=0$ and hence we can express the polynomial in terms of its solutions $x_i$ as $P(x)=2na\prod_{k=1}^{n-1}(x-x_k)$. Equating the two expressions for $P(x)$ we obtain,
\begin{equation}
\prod_{k=1}^{n-1}(x-x_k)=\frac{1}{2na}\left[\prod_{k=1}^{n}(x-y_k+a)-\prod_{k=1}^{n}(x-y_k-a)\right],
\end{equation}
from which relation~(\ref{nice_relation}) follows by taking $x\to y_i+a$.

\subsection{General Considerations}

The application of the simple example presented above allows us to simplify a set of BAE by completely decoupling the $x_i$ fermionic roots, in the case when there are $n-1$ of them and the respective equation has the form~(\ref{simpleBAE}). More commonly it turns out that if our original BAE have $m\le n-1$ $x_i$ roots, though decoupling is no longer possible, we can still replace them in all equations with the remaining $n-1-m$ solutions of the polynomial equation $P(x)=0$, say $\tilde x_i$.

This replacement of the roots $x_i$ with their ``dual'' roots $\tilde x_i$ is only possible for fermionic roots, which don't interact with themselves\footnote{Namely, there exist no $x_i-x_j$ terms in their BAE, which can otherwise be completely general, not just restricted to the simple form (\ref{simpleBAE}).}, allowing all of them to be described by the same single variable equation. The new BAE that will arise from this process will have a different Cartan matrix and Dynkin diagram, and hence fermionic root dualization becomes a method for obtaining BAE corresponding to different choices of Dynkin diagrams, reflecting the existence of this non-unique choice in superalgebras.

Dualization of fermionic roots has been extensively considered for $GL(N|M)$ superalgebras in the context of ${\cal N}=4$
SYM~\cite{Beisert:2005di}, and be can applied with small modifications to our case as well. For a fermionic root which is connected with its adjacent roots with simple lines in the respective Dynkin diagram, such as the $s$ with the $w,r$ roots in the distinguished Dynkin diagram of Figure~\ref{DynkinDiagram}, the corresponding BAE will be
\begin{equation}
\left(\frac{s_i+\frac{i}{2}V_s}{s_i-\frac{i}{2}V_s}\right)^L=\;\:\,\prod_{k=1}^{M_w}\frac{s_i-w_k+i/2}{s_i-w_k-i/2}\prod_{k=1}^{M_r}\frac{s_i-r_k-i/2}{s_i-r_k+i/2},\\
\label{fermionic_BAE}
\end{equation}
where we've now allowed the fermionic root $s$ to also have a spin representation $V_s$ for more generality. This is equivalent to setting the following polynomial to zero,
\begin{equation}
\begin{aligned}
P(s)&=(s+\frac{i}{2}V_s)^L \prod_{k=1}^{M_w}(s-w_k-i/2)\prod_{k=1}^{M_r}(s-r_k+i/2)\\
&\quad-(s-\frac{i}{2}V_s)^L \prod_{k=1}^{M_w}(s-w_k+i/2)\prod_{k=1}^{M_r}(s-r_k-i/2)\\
&=\prod_{k=1}^{M_s}(s-s_k)\prod_{k=1}^{M_{\tilde s}}(s-\tilde s_k),
\end{aligned}
\label{polynomial}
\end{equation}
where $\tilde s_k$ are the dual roots of which there are
\begin{equation}\label{M_tilde_s}
M_{\tilde s}=L+M_w+M_r-M_s-1.
\end{equation}

Similarly with what we did for the simple example, we can now take $s\to w_i\pm i/2$ or $s\to r_i\pm i/2$ and equate the two right-hand sides of (\ref{polynomial}) in order to reexpress the phases
\begin{equation}\label{replace_phases}
\begin{aligned}
\prod_{k=1}^{M_s}\frac{w_i-s_k+i/2}{w_i-s_k-i/2}&=\prod_{k=1}^{M_{\tilde s}}\frac{w_i-\tilde s_k-i/2}{w_i-\tilde s_k+i/2}\prod_{k=1,k\ne i}^{M_w}\frac{w_i-w_k+i}{w_i-w_k-i}\left(\frac{w_i-\frac{i}{2}(V_s-1)}{w_i+\frac{i}{2}(V_s-1)}\right)^L,\\
\prod_{k=1}^{M_s}\frac{r_i-s_k-i/2}{r_i-s_k+i/2}&=\prod_{k=1}^{M_{\tilde s}}\frac{r_i-\tilde s_k+i/2}{r_i-\tilde s_k-i/2}\prod_{k=1,k\ne i}^{M_r}\frac{r_i-r_k-i}{r_i-r_k+i}\left(\frac{r_i-\frac{i}{2}(V_s+1)}{r_i+\frac{i}{2}(V_s+1)}\right)^L,
\end{aligned}
\end{equation}
\FIGURE{
\includegraphics[height=4cm]{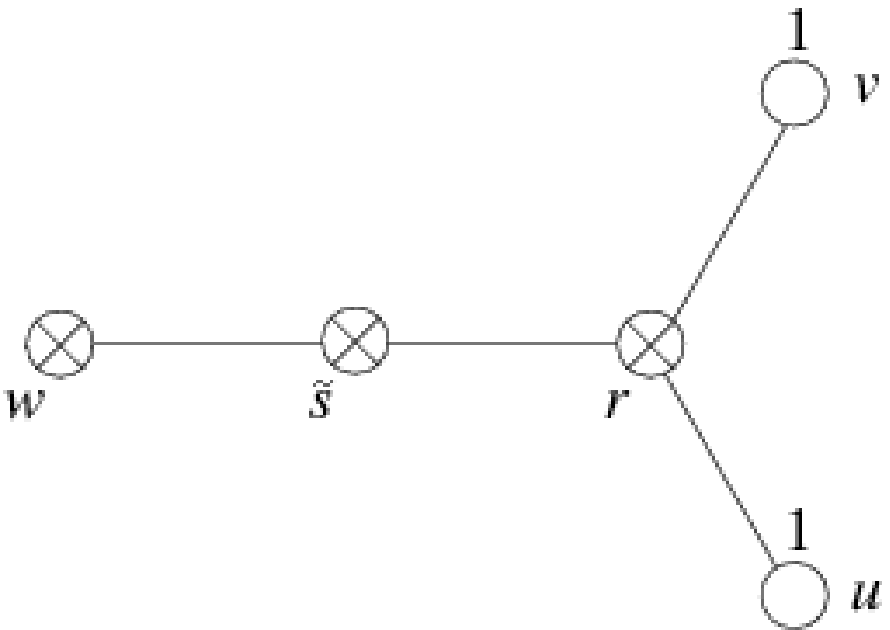}
\caption{The Dynkin diagram that arises from the distinguished one after dualizing the fermionic root $s$.}
\label{dualDynkinD}
}
which are the only factors including the $s_i$ in the remaining Bethe equations (in particular, they appear in the $w$ and $r$ equations), and in this manner eliminate $s_i$ in favor of the $\tilde s_i$\footnote{For the fermionic Bethe equations, we simply replace $s_i\to\tilde s_i$, since they are just different sets of solutions to the same equations.}.The corresponding formulas for the case $V_s=0$  are obtained if we simply set $L\to0$  in (\ref{fermionic_BAE} through (\ref{replace_phases}).
So specifically, if we dualize the $s$ root of the distinguished $OSp(6|4)$ Dynkin diagram, we can see that the $w$ and $r$ roots will also become fermionic because the self-interaction terms in~(\ref{replace_phases}) and~(\ref{betheeq}) cancel out, so the Dynkin diagram corresponding to the new, dualized BAE will be given by that shown in Figure~\ref{dualDynkinD}. Due to~(\ref{M_tilde_s}) and~(\ref{excitations_to_cartan}) the dual root excitation number will be related to the Cartan charges by
\begin{equation}
M_{\tilde s}=\Delta-j-\frac{1}{2}d_1-d_2-\frac{1}{2}d_3-1.
\end{equation}

The analysis becomes only slightly more complicated when we move on to dualize the $r$ root, which is connected to three neighboring roots and has BAE
\begin{equation}\label{dual_rBAE}
1= \prod_{k=1}^{M_s}\frac{r_i-\tilde s_k+i/2}{r_i-\tilde s_k-i/2}\prod_{k=1}^{M_u}\frac{r_i-u_k-i/2}{r_i-u_k+i/2}\prod_{k=1}^{M_v}\frac{r_i-v_k-i/2}{r_i-v_k+i/2}\,.\\
\end{equation}
Because the $u$ and $v$ roots have identical phases, for the purposes of this calculation we can think them as the same root $x$ with a total of $M_x=M_u+M_v$ excitations, $x_i=(u_k,v_l)$. Then (\ref{dual_rBAE}) becomes identical to (\ref{fermionic_BAE}) with $L\to 0, s_i\to r_i, w_i\to \tilde s_i, r_i\to x_i$ and the calculation proceeds as before, this time yielding the BAE corresponding to the Dynkin diagram of Figure~\ref{superD}(a). The excitation number for the $\tilde r$ is expressed in terms of the Cartan charges as
\begin{equation}
M_{\tilde r}=\Delta+L-j-d_1-d_2-d_3-2\,.
\end{equation}
If needed dualization can be performed to the remaining fermionic roots in a similar fashion, for example if we pick one of the two momentum-carrying roots, the only modification~(\ref{fermionic_BAE}) required after we adapt it to the corresponding roots is to replace $i\to 2i$ in the second product. In this manner one can construct different sets of Bethe equations.  Even though their solutions yield the same energies and higher charges, in practice each set is more suitable for studying a certain subset of the spectrum as the equations take a simpler form.

\section{The $OSp(6|4)$ Algebra}\label{osp64_appendix}

Here we mention the form of the $OSp(6|4)$ algebra in a basis where the generators of the maximal compact subgroup $U(2|3)$ are singled out. This particular form is very convenient for constructing representations with the oscillator method~\cite{Gunaydin:1988kz}.

We start with the bosonic spacetime $Sp(4,\mathbb{R})$ algebra,
\begin{equation}
\begin{aligned}
\left[ K_{ij} , P^{kl} \right]
&= \delta_j^l I{}^k{}_i + \delta_i^k I{}^l{}_j + \delta_j^k I{}^l{}_i + \delta_i^l I{}^k{}_j \,,\\
\left[ I{}^i{}_j , P^{kl} \right]
&= \delta_j^k P^{il} + \delta_j^l P^{ik} \,,\\
\left[ I{}^i{}_j , K_{kl} \right]
&= -\delta_k^i K_{jl}-\delta_l^i K_{jk} \,,\\
\left[ I{}^i{}_j , I{}^k{}_l \right]
&= \delta_j^k I{}^i{}_l - \delta_l^i I{}^k{}_j \,,
\end{aligned}
\label{Sp4Ralgebra}
\end{equation}
where we recognize $\Delta = \frac{1}{2} I{}^i{}_i$ as the dilatation operator.

The bosonic R-symmetry algebra $SU(4)$ is given by
\begin{equation}
\begin{aligned}
\left[ A_{\mu\nu} , A^{\rho\sigma} \right]
&= -\delta_\mu^\sigma U{}^\rho{}_\nu + \delta_\mu^\rho U{}^\sigma{}_\nu - \delta_\nu^\rho U{}^\sigma{}_\mu +\delta_\nu^\sigma U{}^\rho{}_\mu \,,\\
\left[ U{}^\mu{}_\nu,A^{\rho \sigma} \right]
&= \delta^\rho_\nu A^{\mu \sigma}+\delta^\sigma_\nu A^{\rho\mu} \,,\\
\left[ U{}^\mu{}_\nu, A_{\rho\sigma} \right]
&= -\delta^\mu_\rho A_{\nu \sigma}-\delta_\sigma^\mu A_{\rho\nu} \,,\\
\left[ U{}^\mu{}_\nu , U{}^\rho{}_\sigma \right]
&= \delta_\nu^\rho U{}^\mu{}_\sigma - \delta_\sigma^\mu U{}^\rho{}_\nu \,.
\end{aligned}
\label{SO6algebra}
\end{equation}

The anticommutators among the odd generators are explicitly given by
\begin{equation}
\label{Osp64_anticommutators}
\begin{aligned}
\{ S_{i\mu} , S^{j\nu} \}       & =\delta_\mu^\nu I{}^j{}_i - \delta_i^j U{}^\nu{}_\mu \,,
&\qquad \{ S_{i\mu} , M{}^j{}_\nu \} &= -\delta_i^j A_{\mu\nu} \,,\\
\{ M{}^i{}_\mu , M{}_j{}^\nu \} & =\delta_\mu^\nu I{}^i{}_j + \delta_j^i U{}^\nu{}_\mu \,,
&\{ S_{i\mu} , M_j{}^\nu \} &= \delta_\mu^\nu K_{ij} \,,
\end{aligned}
\end{equation}
together with others obtained by hermitian conjugation.
Finally, the commutators between even and odd generators are
\begin{equation}
\begin{aligned}
\left[ I{}^i{}_j , M{}^k{}_\mu \right]
&= \delta_j^k M{}^i{}_\mu \,,
& \left[ U{}^\mu{}_\nu , M{}^k{}_\lambda \right]
&= -\delta^\mu_\lambda M{}^k{}_\nu \,,\\
\left[ I{}^i{}_j , M{}_k{}^\mu \right]
&= -\delta^i_k M{}_j{}^\mu\,,
& \left[ U{}^\mu{}_\nu , M{}_k{}^\lambda \right]
&= \delta^\lambda_\nu M{}_k{}^\mu \,,\\
\left[ I{}^i{}_j , S_{k\mu} \right]
&= -\delta^i_k S_{j\mu} \,,
& \left[ U{}^\mu{}_\nu , S_{k\lambda} \right]
&= -\delta_\lambda^\mu S_{k\nu} \,,\\
\left[ I{}^i{}_j , S^{k\mu} \right]
&= \delta_j^k S^{i\mu} \,,
& \left[ U{}^\mu{}_\nu , S^{k\lambda} \right]
&= \delta^\lambda_\nu S^{k\mu} \,,\\
\left[ K_{ij} , M{}^k{}_\mu \right]
&= \delta_i^k S_{j\mu} + \delta_j^k S_{i\mu} \,,
&\qquad \left[ A_{\mu\nu} , M{}^\lambda{}_k \right]
&= -\delta^\lambda_\mu S_{\nu k} + \delta^\lambda_\nu S_{\mu k} \,,\\
\left[ K_{ij} , S^{k\mu} \right]
&= \delta_i^k M{}_j{}^\mu + \delta_j^k M{}_i{}^\mu \,,
& \left[ A_{\mu\nu} , S^{k\lambda} \right]
&= -\delta^\lambda_\mu M{}^k{}_\nu + \delta^\lambda_\nu M{}^k{}_\mu \,,
\end{aligned}
\label{OSp64algebra}
\end{equation}
where we have again omitted commutators which can be obtained from these
by hermitian conjugation.

As we mentioned in the beginning of this appendix, the advantage of this basis is that it makes the mapping of generators to bilinears of superoscillators very straightforward. In particular, for a length-$f$ representation we need $f=2p+\epsilon$ $U(2|3)$ contravariant and covariant superoscillators, where $p=[\frac{f}{2}]$ and $\epsilon$ is either zero or one, defined as
\begin{equation}
\begin{aligned}
\xi_A(r) &= \begin{pmatrix} a_i(r) \cr \alpha_\mu(r) \end{pmatrix} \,, & \qquad \xi^A(r) &= \xi_A(r)^\dag = \begin{pmatrix} a^i(r) \cr \alpha^\mu(r) \end{pmatrix} = \syng(1) \,,\\
\eta_A(r) &= \begin{pmatrix} b_i(r) \cr \beta_\mu(r) \end{pmatrix} \,, & \eta^A(r) &= \eta_A(r)^\dag = \begin{pmatrix} b^i(r) \cr \beta^\mu(r) \end{pmatrix} = \syng(1) \,,\\
\zeta_A & = \begin{pmatrix} c_i \cr \gamma_\mu \end{pmatrix} \,, & \zeta^A &= {\zeta_A}^\dag = \begin{pmatrix} c^i \cr \gamma^\mu \end{pmatrix} = \syng(1) \,,
\end{aligned}
\label{superoscillators}
\end{equation}
with the super-index
$A$ taking the values $1,2|1,2,3$ and $r=1,\dots,p$.

Then all the $OSp(6|4)$ generators we presented above are written in terms of these superoscillators as
\begin{equation}
\begin{aligned}
S^{AB}    &= \vec{\xi}^A \cdot \vec{\eta}^B + \vec{\eta}^A \cdot \vec{\xi}^B + \epsilon \; \zeta^A \zeta^B
&&= \syng(2) \,,\\
S_{AB}    &= \vec{\xi}_A \cdot \vec{\eta}_B + \vec{\eta}_A \cdot \vec{\xi}_B + \epsilon \; \zeta_A \zeta_B
&&= (S^{BA})^\dagger \,,\\
M{}^A{}_B &= \vec{\xi}^A \cdot \vec{\xi}_B + (-1)^{(\deg A)(\deg B)} \vec{\eta}_B \cdot \vec{\eta}^A \\
&\quad + \frac{\epsilon}{2} \left( \zeta^A \zeta_B + (-1)^{(\deg A)(\deg B)} \zeta_B \zeta^A \right)
&&= (M{}^B{}_A)^\dagger \,,
\end{aligned}
\label{OSp64Rgenerators}
\end{equation}
where in order to avoid confusion when the superindices take specific values we have renamed
\begin{equation}
\begin{aligned}
M{}^i{}_j \to I{}^i{}_j && M{}^\mu{}_\nu \to U{}^\mu{}_\nu&&M{}^\mu{}_i\to M{}_i{}^{\mu}\\
S_{ij} \to K_{ij}&&S_{\mu\nu} \to A_{\mu\nu}\\
S^{ij} \to P^{ij}&& S^{\mu\nu} \to A^{\mu\nu}
\end{aligned}
\end{equation}

\end{document}